\documentclass[12pt]{article}
\usepackage{amssymb,amsmath}
\usepackage{epsfig}
\usepackage{epstopdf}
\usepackage{latexsym}
\usepackage{graphicx}
\usepackage{subfigure}
\usepackage{color}
\usepackage{datetime}
\usepackage[
      colorlinks=false,
      linkcolor=blue,  
      urlcolor=blue,    
      filecolor=blue,
      citecolor=red,
      pdfstartview=FitV,
      linktocpage=true,
       bookmarksopen=true    
      ]{hyperref}
\def\oneone{\rlap 1\mkern4mu{\rm l}}
\def\coeff#1#2{\relax{\textstyle {#1 \over #2}}\displaystyle}
\def\ds{\displaystyle}

\def\IR{\mathbb{R}}
\def\ZZ{\mathbb{Z}}

\def\cM{{\cal M}}
\def\cN{{\cal N}}

\def\cS{{\cal S}}

\def\cX{{\cal X}}
\def\Neql#1{{\cal N}\!=\!{#1}}
\definecolor{cardinal}{rgb}{0.6,0,0}
\definecolor{darkgreen}{rgb}{0,0.5,0}
\definecolor{golden}{rgb}{0.92, 0.7, 0}
\definecolor{midnight}{rgb}{0, 0, 0.5}
\definecolor{darkblue}{rgb}{0.2, 0, 0.8}

\begin{document}  
%%%%%%%%%%%%%%%%%%%%%%%%%%%%%%

\begin{titlepage}
 
\bigskip
\bigskip
\bigskip
\bigskip
\centerline{\Large \bf Supersymmetric Charged Clouds in $AdS_5$}
\bigskip
\bigskip
\centerline{{\bf Nikolay Bobev,  Arnab Kundu,}}
\centerline{{\bf Krzysztof Pilch  and Nicholas P. Warner }}
\bigskip
\centerline{Department of Physics and Astronomy}
\centerline{University of Southern California} \centerline{Los
Angeles, CA 90089, USA}
\bigskip
\centerline{{\rm bobev@usc.edu,~ akundu@usc.edu,~pilch@usc.edu,~warner@usc.edu } }
\bigskip
\bigskip

\begin{abstract}

\noindent  We consider supersymmetric holographic flows that involve background gauge fields dual to chemical potentials in the boundary field theory.
We use a consistent truncation of gauged $\Neql8$ supergravity in five dimensions and we give a complete analysis of the supersymmetry conditions for a large family of flows.
We examine how the well-known supersymmetric flow between two fixed points is  modified by the presence of the chemical potentials and this yields a new, completely smooth, solution that interpolates between two global $AdS$ spaces of different radii and with different values of the chemical potential.  We also examine some black-hole-like singular flows and a new non-supersymmetric black hole solution.  We comment on the interpretation of our new solutions in terms of giant gravitons and discuss the implications of our work for finding black-hole solutions in $AdS$ geometries.

\end{abstract}

\end{titlepage}

%%%%%%%%%%%%%%%%%%%%%%%%%%%%%%%%%%%%%

\tableofcontents

%%%%%%%%%%%%%%%%%%%%%%%%%%%%%%%%%%%%%
\section{Introduction}
%%%%%%%%%%%%%%%%%%%%%%%%%%%%%%%%%%%%%

The application of holographic field theory to condensed matter systems has generated new interest in studying richer families of holographic RG flows \cite{Hartnoll:2009sz,Herzog:2009xv}. In particular, there has been a focus on systems in which the Poincar\'e symmetry is broken down to some form of Galilean symmetry or, perhaps, a Schr\"odinger symmetry.  Moreover, in the study of holographic superconductors,\footnote{See \cite{Horowitz:2010gk} for recent review and a comprehensive list of references.} one of the goals is to induce a condensate and this is generically done through chemical potentials whose holographic duals are electrostatic fields in the bulk.  One is thus led to the study of holographic flows in the presence of, at least, Coulomb potentials and in which the spatial and temporal parts of the metric receive different ``warp factors.''  One of the purposes of this paper is to re-examine supersymmetric holographic flows in this context and see to what extent the supersymmetric solutions might play a r\^ole in AdS/CMT. Since our solutions will be obtained from a consistent truncation of $\mathcal{N}=8$ five-dimensional supergravity they are also clearly of interest as dual to phases of four-dimensional $\mathcal{N}=4$ SYM at finite chemical potential.  

The holographic flows we consider involve setting up a space-time that is asymptotically $AdS$ with background electromagnetic fields. The study of such holographic flow solutions will also provide a new perspective on the study of black holes in anti-de Sitter geometries.  In particular, one might ultimately find interesting new black-hole-like solutions in $AdS$ that consist of a black hole in the center and a charged cloud surrounding the black hole so that the charges at infinity are a non-trivial combination of those of the black hole and the charged cloud.   Once again we will focus on supersymmetric configurations because of simplicity and stability and we will find a supersymmetric charged cloud in global $AdS$ that spontaneously breaks an Abelian gauge symmetry in the gravitational background. That is, we find a charged solution that interpolates between global $AdS$ with one set of electrostatic potentials  at infinity and global $AdS$ with a different radius and different electrostatic potentials in the core.  The voltage difference is accounted for by a charged cloud distributed in a region whose scale is set by the curvature of the $S^3$  in the global $AdS$.  We find that the charge density of this cloud vanishes extremely fast in the core and so there is no obvious obstruction to putting a black hole (with an orthogonal set of charges) in the middle of this cloud.  We make some first steps in this direction by examining simple black-hole solutions that might be placed inside such a charged cloud.

It is also important to recall that turning on Coulomb fields in four or five dimensions corresponds to turning on angular momenta on the internal manifold that is used to compactify the underlying M-theory or IIB supergravity \cite{Cvetic:1999xp}. Thus all of the solutions one finds in this manner will have potentially interesting brane interpretations as some form of giant gravitons (or superstars) \cite{McGreevy:2000cw,Myers:2001aq}.  Indeed, even though some of the black-hole-like solutions we find are rather singular in five dimensions, based on the criteria of \cite{Gubser:2000nd}, their uplifts can almost certainly be given some natural brane interpretation.  

Another interesting aspect of the work presented here is that we are led to consider supersymmetric flows that start, in the UV,  from either Poincar\'e or global $AdS$ geometries.  That is, we consider field theories on both $\IR^{3,1}$ and $S^3 \times \IR$.    Indeed, it seems to be very natural to combine field theories dual to global $AdS$ with chemical potentials:  The finite volume lifts the ground state degeneracy of the field theory and, in particular,  conformal invariance requires the field-theory scalars to have conformal couplings which, on $S^3$, behave like a mass.  On the other hand, the presence of a chemical potential induces an instability, essentially a negative mass-squared term, which, in AdS/CMT, usually generates the condensate.  In global $AdS$ there is thus a competition between these two contributions to the mass terms and we find that, while there are infinite families of supersymmetric flows for field theories on both $\IR^{3,1}$ and $S^3 \times \IR$, the only smooth supersymmetric flows to non-trivial fixed points are the well-known one with a Poincar\'e invariant space-time \cite{Freedman:1999gp}  and a new one on  $S^3 \times \IR$ with the chemical potentials precisely tuned in terms of the curvature of the $S^3$.  The new flow is asymptotic to global $AdS_5$ in both the UV and IR, but with different radii and  different electrostatic potentials.  In between the IR and UV there is a charged, supersymmetric cloud that gives rise to this potential difference. 

One of the issues in the construction of holographic superconductors is to embed the symmetry breaking gravity solution in string theory (so that one has a tried-and-true holographic dictionary). This was the approach followed in \cite{Gubser:2009qm,Gauntlett:2009dn,Gubser:2009gp,Gauntlett:2009bh} where the authors successfully embedded a model of a holographic superconductor in IIB and eleven-dimensional supergravity.  However the end point of the zero-temperature, symmetry breaking domain walls of \cite{Gubser:2009qm,Gauntlett:2009dn,Gubser:2009gp,Gauntlett:2009bh} are non-supersymmetric $AdS$ solutions, which for compactifications based on spheres are known to be unstable \cite{Pilch,NickH}. The easiest way to ensure stability is to find a flow that is supersymmetric and this was one of the original motivations for this work.  The obvious  objection to this idea is that the formation of condensate cannot be supersymmetric because it is evidently not a ground-state of the original Hamiltonian.  However, the condensate is necessarily charged and it might well saturate a BPS bound and thereby preserve supersymmetry.  Indeed, the supersymmetric solutions that we find are BPS for this reason.  However, the flows to the non-trivial supersymmetric fixed point do not involve a condensate: just as in  \cite{Freedman:1999gp}  these flows involve turning on a supersymmetric mass term that breaks the supersymmetry from $\Neql4$ to $\Neql1$.   We also exhibit supersymmetric flows that involve chemical potentials and the fermion condensate,  but these do not go to a smooth fixed-point solution but have some kind of black-hole-like naked singularity in the core.

While this work has natural generalizations to  M-theory and holographic field theories in $(2+1)$ dimensions, this paper will focus on the IIB theory and  $(3+1)$-dimensional holographic field theories.  More specifically, we will work with gauged  $\Neql8$ supergravity in five dimensions, and use a consistent truncation, defined in  \cite{Khavaev:2000gb}, that reduces it to an Abelian gauged $\Neql2$ supergravity theory, in five dimensions, coupled to two vector multiplets and four differently charged hypermultiplets.  We further reduce the scalar sector so that we can focus on holographic flows that involve fundamental bilinear operators in the bosons and fermions of the dual field theory.    This enables us to define a large class of $\Neql1$ supersymmetric flows that involve chemical potentials, masses and vevs in the dual $\Neql1$ Yang-Mills theory.

In section 2 we define the  $\Neql2$ supergravity theory as a consistent truncation of the  $\Neql8$ theory and then perform a further (consistent) truncation to the sector of interest.  We then give the field theory action and superpotential.  Section 3 contains a detailed analysis of the supersymmetry conditions for the holographic flows involving all the scalars and Coulomb fields.  Section 4 contains the details of the new smooth flow between the maximally symmetric fixed point and the non-trivial supersymmetric fixed point of \cite{Khavaev:1998fb}.  As described above, this new flow involves non-trivial chemical potentials and runs between two {\it global} $AdS$ spaces.  In Section 5 we consider some supersymmetric black-hole-like singular flows and show that our results here match some of the known solutions in the literature.  We also discuss a new solution that consists of a non-supersymmetric black hole in the background of the non-trivial supersymmetric fixed point.  In section 6 we give the conclusions and describe several promising avenues for further research.

%%%%%%%%%%%%%%%%%%%%%%%%%%%%%%%%%%%%%
\section{Truncating the gauged $\Neql8$ supergravity theory}
%%%%%%%%%%%%%%%%%%%%%%%%%%%%%%%%%%%%%

The forty-two scalars of  $\Neql8$ supergravity parametrize the non-compact coset
space $E_{6(6)}/USp(8)$ and it is convenient to write the  $E_{6(6)}$  Lie algebra in terms of a real $27 \times 27$ matrix in the 
 $SL(6,\IR)\times SL(2,\IR)$ basis \cite{Gunaydin:1985cu}:
\begin{equation}
\cX ~=~ \left( 
\begin{array}{cc}
-4 \, \Lambda^{[M}{}_{[I}\delta^{N]}{}_{J]} & \sqrt{2} \, \Sigma_{IJP\beta} \\
\sqrt{2} \, \Sigma^{MNK\alpha} & {\Lambda^K}_P\, \delta^{\alpha}{}_{\beta} \, + \, {\Lambda^{\alpha}}_{\beta}\, \delta^{K}{}_{P} 
\end{array}\right) \,,
\end{equation}
where $I =1,\dots, 6$; $\alpha = 1,2$; the matrices ${\Lambda^K}_P$, ${\Lambda^\alpha}_\beta$  represent
elements of the $SL(6,\IR)\times SL(2,\IR)$ Lie algebra and $\Sigma_{IJP\beta}$ transforms in the ${\bf (20,2)}$ of
$SL(6,\IR)\times SL(2,\IR)$.  Raising of the indices on $\Sigma$ is done using the $\epsilon$-symbols.

The truncation to the $\Neql2$ supergravity theory is then obtained by finding the sector of gauged $\Neql{8}$  supergravity that commutes with the 
 $\ZZ_2 \times \ZZ_2$ subgroup of $SO(6)$ defined by the matrices:
\begin{equation}
{\rm diag} (-1, -1, -1, -1, +1, +1) \quad \hbox{ \rm and } \quad {\rm diag} (+1, +1, -1, -1, -1, -1)  \,.
  \label{Ztwos}
\end{equation} 
One can easily verify that the corresponding $SU(4)$ matrices acting on the ${\bf 4}$ or ${\bf \bar 4}$ are: 
\begin{equation}
{\rm diag} (-1,  +1, -1, +1)\, \quad \hbox{ \rm and } \quad  {\rm diag} (-1, -1,  +1, +1)   \,.
  \label{spinZtwos}
\end{equation} 
There is thus one invariant spinor in each of the ${\bf 4}$ and ${\bf \bar 4}$, which means that the truncation is indeed $\Neql{2}$ supersymmetric.

The tensor gauge fields transform in the $\mathbf{6}$ of $SO(6)$ and so none of them is invariant under (\ref{Ztwos}).   There are three vector fields, $A^{(1)}_{\mu} \equiv A_\mu^{12}$,  $A^{(2)}_{\mu} \equiv A_\mu^{34}$ and  $A^{(3)}_{\mu}  \equiv A_\mu^{56}$, that are invariant under  (\ref{Ztwos}) and one of these must be the graviphoton.  This means that the $\Neql2$ supergravity must be coupled to two vector muliplets.

Invariance under the $\ZZ_2 \times \ZZ_2$ in   (\ref{Ztwos}) breaks the $SL(6,\IR)$  to the block  diagonal $S((GL(2,\IR))^3) =   (SL(2,\IR))^3 \times (SO(1,1))^2$.  The $SO(1,1)$ factors are multiples of the $2 \times 2$ identity matrices and can be parametrized by
\begin{equation}
\cS  ~\equiv~ {\rm diag} ( -\alpha+\beta, -\alpha+ \beta, -\alpha -\beta, -\alpha -\beta,~  2\,\alpha,~ 2\, \alpha\, )\,.
  \label{metdefm}
\end{equation} 
These two scalar fields belong to the vector multiplets\footnote{Note that we are reversing the sign $\alpha \to -\alpha$ compared to the conventions of   \cite{Khavaev:2000gb}  so as to bring the conventions of this paper into line with earlier work, such as \cite{Freedman:1999gp}.}.

Invariant tensors of the form,   $\Sigma_{IJP\beta}$,  are only non-zero if  the three indices $I,J,K$ lie in some permutation of the three distinct index sets, $\{1,2\}, \{3,4\}, \{5,6\}$.  There are thus $16$ such generators transforming in the  $(\mathbf{2},\mathbf{2},\mathbf{2},\mathbf{2})$ of  $(SL(2,\IR))^4$, where the last  $SL(2,\IR)$ is that of the dilaton and axion of type IIB supergravity.  In these $16$ generators there are $8$ anti-self-dual forms and $8$  self-dual forms, which correspond, respectively, to compact and non-compact $E_{6(6)}$ generators.  These forms extend the $(SL(2,\IR))^4$ to $SO(4,4)$ and thus the scalar coset of the $\Neql2$ supergravity theory is a quaternionic K\"ahler manifold with extra $SO(1,1)$ factors:
\begin{equation}
\cM = \mathcal{M}_{QK} \times \mathcal{M}_{VS}  ~\equiv~ {SO(4,4) \over  SO(4) \times SO(4)} \times \left(SO(1,1)  \times SO(1,1)\right)\,.
\label{scalmanifold}
\end{equation} 
This scalar manifold, of course, fits with the general classification of matter coupled to $\mathcal{N}=2$ gauged supergravity in five dimensions \cite{Ceresole:2000jd}. The scalar manifold is a product of a quaternionic K\"ahler, $\mathcal{M}_{QK}$, and a very special, $\mathcal{M}_{VS}$, manifold. The scalars in the quaternionic K\"ahler manifold lie in four charged hypermultiplets.  
  
One can make further consistent truncations of this theory.  Most particularly, one can truncate the hypermultiplet sector in a variety of ways.\footnote{It was shown in \cite{Ceresole:2001wi} that setting $\beta=0$ and keeping only one of the hypermultiplets yields a consistent truncation of our model with a scalar manifold  ${SU(2,1) \over  SU(2) \times U(1)} \times SO(1,1)$. This subsector is the $SU(2)\times U(1)$ invariant truncation and contains the supersymmetric flow of \cite{Freedman:1999gp}. }   For example, in \cite{Liu:2007rv} the truncation to the ``incomplete hypermutiplets'' parametrized by  $(SL(2,\IR))^4 \times (SO(1,1))^2 \subset SL(6,\IR)   \times SL(2,\IR)$ was considered.  This truncation was thus restricted to metric modes and the dilaton and axion in the IIB theory.  In the holographic dual, this essentially restricts to bilinears of the bosonic fields.   The truncation considered in 
 \cite{Khavaev:2000gb} was almost the orthogonal sector to that considered in \cite{Liu:2007rv}.  The coset is, once again, 
\begin{equation}
\cM ~=~ \bigg({SU(1,1) \over U(1)} \bigg)^4 ~ \times~ SO(1,1) ~\times~  SO(1,1) \,.
  \label{coset}
\end{equation} 
The $SO(1,1)$ factors are simply those of (\ref{metdefm}) and correspond to scalars in the vector multiplets.  The non-compact generators of $SU(1,1) \cong SL(2,\IR)$   are now those provided by the $\ZZ_2 \times \ZZ_2$ invariant self-dual forms $\Sigma_{IJP\beta}$.   In the IIB theory these correspond to fluxes and in the holographic dual theory they correspond to {\it fermion} bilinears.   This particular truncation was also shown to be consistent in  \cite{Khavaev:2000gb}  because it is the invariant subsector under a further $\ZZ_4$ symmetry.

The $SU(1,1) /U(1)$ cosets can be naturally described in terms of complex scalars, $\zeta_j$, $j=1, \dots,4$ and we will parametrize these in terms of a magnitude and a phase,  with $\zeta_j = \tanh(\varphi_j) e^{i \theta_j}$.  The Lagrangian of this subsector is given by truncating the action in \cite{Gunaydin:1985cu}\footnote{Throughout this paper, our conventions will be precisely those of \cite{Gunaydin:1985cu}.}:
\begin{equation}
\begin{split}
e^{-1} {\cal L} &=   -\coeff{1}{4}\,R ~-~ \coeff{1}{4}\,  \Big[ \rho^{4}   \nu^{-4}  F^{(1)}_{\mu \nu} F^{(1)\, \mu \nu}  ~+~ \rho^{4}   \nu^{4}  F^{(2)}_{\mu \nu} F^{(2)\, \mu \nu} ~+~ \rho^{-8}   F^{(3)}_{\mu \nu} F^{(3)\, \mu \nu}   \Big]  \\ 
 &+~  \coeff{1}{2}\, \sum_{j=1}^4 \,(\partial_\mu \varphi_j)^2 ~+~3 (\partial_\mu \alpha)^2 ~+~(\partial_\mu \beta)^2  \\
 &+~  \coeff{1}{8}\, \sinh^2 (2\varphi_1) \big(\partial_\mu \theta_1 + (A^{(1)}_\mu+ A^{(2)}_\mu -A^{(3)}_\mu) \big)^2   \\[7pt] 
 &+~\coeff{1}{8}\, \sinh^2 (2\varphi_2) \big(\partial_\mu \theta_2 + (A^{(1)}_\mu-A^{(2)}_\mu +A^{(3)}_\mu) \big)^2   \\[7pt]
 &+~\coeff{1}{8}\, \sinh^2 (2\varphi_3) \big(\partial_\mu \theta_3 + (- A^{(1)}_\mu+ A^{(2)}_\mu +A^{(3)}_\mu) \big)^2  \\[7pt]
 &+~\coeff{1}{8}\, \sinh^2 (2\varphi_4) \big(\partial_\mu \theta_4 - (A^{(1)}_\mu+ A^{(2)}_\mu +A^{(3)}_\mu) \big)^2   ~-~ {\cal P}   \,,
\label{Lagrangian}
\end{split}
\end{equation} 
where the $F^{(J)}$ are the field strengths of the $U(1)$ gauge fields, $A^{(J)}$, and ${\cal P}$ is the scalar potential.  
We have also the exponentiated matrix elements of the $SO(1,1)$ factors:
\begin{equation}
\rho~\equiv~e^\alpha\,, \qquad \nu ~\equiv~ e^\beta \,.
  \label{rhonudefn}
\end{equation} 
In  \cite{Khavaev:2000gb}  it was shown that the scalar potential,  ${\cal P}$, is given in terms of a superpotential:
\begin{equation}
 {\cal P} ~=~{g^2\over 8}\,  \bigg[\,  \sum_{j=1}^4 \, \bigg({\partial W \over \partial \varphi_j}\bigg)^2 ~+~  {1\over 6}\,     \bigg({\partial W \over \partial \alpha }\bigg)^2 ~+~  {1\over 2}\,     \bigg({\partial W \over \partial \beta }\bigg)^2\, \bigg]  ~-~  {g^2 \over 3}\,    W^2 \,,
\label{PsuperP}
\end{equation} 
where
\begin{eqnarray}
W &=& - {1 \over 4\rho^2\nu^2}\, \big[\,   (1 + \nu^4  - \nu^2 \rho^6  ) \,
\cosh(2\, \varphi_1) ~+~(- 1 + \nu^4  + \nu^2 \rho^6  ) \,  \cosh(2\, \varphi_2)\notag\\ 
&& \qquad \quad ~+~  (1 -  \nu^4  + \nu^2 \rho^6  \big) \,  
\cosh(2\, \varphi_3) ~+~(1 + \nu^4  + \nu^2 \rho^6  \big) \,\cosh(2\, \varphi_4) \,\big] \,.
  \label{Wdefn}
\end{eqnarray} 

One should also note that $W$ is {\it invariant, up to a sign,} under the
permutation group $S_3$.  These permutations are generated
by the transformations:
\begin{eqnarray}
 p_1 ~:~  &&   \varphi_1 \leftrightarrow
\varphi_3 \,, \quad \varphi_2 \to \varphi_2 \,,  \ \quad \varphi_4 \to \varphi_4 \,,  \quad
\alpha \to  \coeff{1}{2}(\beta-\alpha) \,, \quad \beta \to
\coeff{1}{2}(\beta + 3\,\alpha)\,, \nonumber \\
p_2 ~:~ &&    \varphi_2 \leftrightarrow
\varphi_3 \,, \quad \varphi_1 \to \varphi_1 \,, \quad \varphi_4 \to \varphi_4~, \quad
\alpha \to \alpha\,, \quad \beta \to -\beta\,,
\label{perms}
\end{eqnarray} 
and these act on $W$ according to: $p_1:W \to  W$ and $p_2:W \to W$.  

In the dual $\mathcal{N}=4$ Yang-Mills theory, the neutral supergravity scalars are dual do the boson bilinears:  
\begin{eqnarray}
\alpha &\longleftrightarrow&  -\text{Tr}( X_1^2 +X_2^2+X_3^2 + X_4^2 - 2 X_5^2 - 2X_6^2)~, \\
\beta &\longleftrightarrow& \text{Tr}( X_1^2 +X_2^2 - X_3^2 - X_4^2 )~,
\label{abduals}
\end{eqnarray} 
while the supergravity charged scalars are dual to fermion bilinears:
\begin{equation}
\varphi_j ~\longleftrightarrow~ \text{Tr}( \lambda_j\lambda_j) +\text{h.c.} \,.
\label{phiduals}
\end{equation} 
Here $X_a$ and $\lambda_j$ are the usual 6 scalars and 4 fermions of $\mathcal{N}=4$ Yang-Mills theory. In the neighborhood of the $SO(6)$ critical point, $\alpha = \beta = \varphi_j =0$, flows induced by the superpotential,   (\ref{Wdefn}), result in non-normalizable modes  for small $\varphi_1$, $\varphi_2$, $\varphi_3$ and normalizable modes for $\varphi_4$.  Thus a generic flow from the maximally supersymmetric point involves masses for the fermions $\lambda_1, \lambda_2, \lambda_3$ and a condensate for $\lambda_4$.

It is worth noting that if we set $\beta=\varphi_2=\varphi_3=0$ we get a further consistent truncation to the $SU(2)\times U(1)$ invariant sector of five-dimensional gauged supergravity. In the notation of  \cite{Freedman:1999gp} this is given by
\begin{equation}
\phi^{FGPW} = 0~, \qquad\qquad \alpha^{FGPW} = \alpha~, \qquad\qquad \varphi_1^{FGPW} = \varphi_1~, \qquad\qquad \varphi_2^{FGPW} = \varphi_4~.
\label{FGPWscalars}
\end{equation} 

The potential \eqref{PsuperP} has four critical points which correspond to $AdS_5$ vacua in the $SU(2)\times U(1)$ invariant sector of five-dimensional gauged supergravity\footnote{Note that the $SO(5)$ invariant critical point of \cite{Khavaev:1998fb} is in the $SU(2)$ and not in the $SU(2)\times U(1)$ invariant sector of five-dimensional gauged supergravity and therefore is not a critical point of the potential \eqref{PsuperP}.} \cite{Khavaev:1998fb}:

\begin{itemize}

\item The $SO(6)$ point with $\mathcal{N}=8$ supersymmetry:
\begin{equation}
\alpha=\beta=\varphi_j=0~, \qquad\qquad  \mathcal{P} = - \ds\frac{3}{4} g^2 ~.
\label{SO6point}
\end{equation} 

\item The $SU(2)\times U(1)$ point with $\mathcal{N}=2$ supersymmetry:
\begin{equation}
\alpha= \ds\frac{1}{6} \log 2~, \qquad \varphi_1= \ds\frac{1}{2} \log 3~,  \qquad \beta=\varphi_2=\varphi_3=\varphi_4=0~, \qquad  \mathcal{P} = - \ds\frac{2^{1/3} 2 }{3} g^2 ~.
\label{PWpoint}
\end{equation} 

\item The non-supersymmetric $SU(3)$ point:
\begin{equation}
\alpha= 0~, \qquad \varphi_1= \ds\frac{1}{2} \log (2-\sqrt{3})~, \qquad\beta=\varphi_2=\varphi_3=\varphi_4=0~,  \qquad  \mathcal{P} = - \ds\frac{27}{32} g^2 ~.
\label{SU3point}
\end{equation} 

\item The non-supersymmetric $SU(2) \times U(1) \times U(1)$ point:
\begin{equation}
\begin{split}
\alpha &= \ds\frac{1}{12} \log 10, \qquad \varphi_1=\varphi_4 = \ds\frac{1}{4} \log\left( \ds\frac{11-4\sqrt{6}}{5} \right)\,, \qquad \beta=\varphi_2=\varphi_3=0\,,\\  
\mathcal{P} &= - \ds\frac{3\, 5^{2/3} }{2^{10/3}} g^2~.
\label{SU2U1U1point}
\end{split}
\end{equation} 
\end{itemize}

The first two fixed points above are supersymmetric and therefore stable, while the last two fixed points are known to be unstable \cite{Pilch,Girardello:1999bd}. The non-supersymmetric $SU(3)$ point uplifts to Romans' solution in IIB supergravity \cite{Romans:1984an} and is the end point of the non-supersymmetric domain wall solutions studied in \cite{Girardello:1998pd,Distler:1998gb,Gubser:2009gp}. 

%%%%%%%%%%%%%%%%%%%%%%%%%%%%%%%%%%%%%
\section{The supersymmetric flows}
%%%%%%%%%%%%%%%%%%%%%%%%%%%%%%%%%%%%%

 %%%%%%%%%%%%%%%%%%% 
\subsection{The background}
%%%%%%%%%%%%%%%%%%% 
 
We want to consider systems with electrostatic charges and this breaks Lorentz invarince, so that the time and space components of the metric will have distinct warp factors.  In particular, we want to consider solutions in which there is a radial coordinate upon which everything depends, but are otherwise maximally symmetric in the spatial directions.  This is a common situation in  
many of the proposed holographic duals of condensed matter systems.  Thus we take the metric to be either of the form:
\begin{equation}
ds_5^2 ~=~ e^{2\, A(r)} \,\big[  f(r)^2 dt^2 ~-~ d\vec x \cdot d \vec x \, \big]  ~-~ {dr^2 \over  f(r)^2} \,,
  \label{fivemet1}
\end{equation} 
or of the form
\begin{equation}
ds_5^2 ~=~ e^{2\, A(r)} \,\big[  f(r)^2 dt^2 ~-~ \coeff{1}{4} a^2 (\sigma_1^2 + \sigma_2^2 + \sigma_3^2)  \, \big]  ~-~ {dr^2 \over f(r)^2} \,,
  \label{fivemet2}
\end{equation} 
where $a$ is a constant parameter.  The $\sigma_j$ are the $SU(2)$ left-invariant $1$-forms
\begin{eqnarray}\label{eqt: sigma}
\sigma_1 &=& \cos\alpha_3 ~ d\alpha_1 + \sin\alpha_1 ~\sin\alpha_3
~d\alpha_2~,\notag\\
\sigma_2 &=& \sin\alpha_3 ~ d\alpha_1 - \sin\alpha_1 ~\cos\alpha_3
~d\alpha_2~,\label{sigmas}\\
\sigma_3 &=& d\alpha_3 + \cos\alpha_1 ~d\alpha_2~,\notag
\end{eqnarray}
which satisfy $d \sigma_i = {1 \over 2} \epsilon_{ijk} \sigma_j \wedge \sigma_k$. The metric on the unit radius $S^3$ is
\begin{equation}
ds^2_{S^3} =  \coeff{1}{4} (\sigma_1^2 + \sigma_2^2 + \sigma_3^2)~.
\end{equation}
The metric on $AdS_5$ of radius $L$ in this set of coordinates is:
\begin{equation}
ds^2_{AdS_5} ~=~ e^{2r/L}\left(\left( 1+ \ds\frac{L^2}{a^2}e^{-2r/L} \right) dt^2 - \coeff{1}{4} a^2 (\sigma_1^2 + \sigma_2^2 + \sigma_3^2)\right) - \left( 1+ \ds\frac{L^2}{a^2}e^{-2r/L} \right)^{-1} dr^2~.
\label{AdSflowcoords}
\end{equation}
We are thus going to consider solutions  that are either asymptotic to Poincar\'e $AdS_5$ and based upon (\ref{fivemet1}), or solutions that are asymptotic to global $AdS_5$ and based upon (\ref{fivemet2}).  As one should expect, the equations governing these two classes of solutions are very similar and one can obtain the equations for the metric  (\ref{fivemet1}) by taking the radius, $a$, of the $S^3$ in  (\ref{fivemet2}) to be infinite.

We will also adopt the obvious sets of frames:
\begin{equation}
e^0 ~=~ e^{A} \,  f \, dt \,, \quad e^i ~=~ e^{A} \,  dx^i \ \ {\rm or}  \ \ e^i ~=~ \coeff{a}{2}\, e^{A} \,  \sigma_i  \,, \quad e^4 ~=~   f^{-1} \, dr \,,
  \label{frames1}
\end{equation} 
and the gamma matrix conventions of \cite{Gunaydin:1985cu}, with:
\begin{equation}
\big\{ \gamma^a\,, \gamma^b\big\} ~=~   2\, \eta^{ab} \,,
  \label{gammconv1}
\end{equation} 
where $\eta = {\rm diag}(+1,-1,-1,-1,-1)$.  In particular, we take: 
\begin{equation}
\gamma^0 \gamma^1 \gamma^2 \gamma^3 \gamma^4 ~=~   \oneone \,.
  \label{gammconv2}
\end{equation} 

In this paper we are going to focus on solutions with electric charges and so we take the Maxwell fields to be:
\begin{equation}
A^{(I)} ~=~  \Phi_I (r) \,dt \,,  \qquad  F^{(I)} ~=~ -(\partial_r\Phi_I ) \, dt \wedge dr \qquad I =1,2,3 \,.
  \label{Maxfields1}
\end{equation} 
We will also only seek solutions in which the phases, $\theta_j$, of the scalar fields are constant.   This means that we have fixed a gauge for the $U(1)$ gauge fields.  The ability to shift the potentials by an overall constant is, of course, equivalent to allowing the $\theta_j$ to have a linear dependence on $t$. To maintain the constancy of the $\theta_j$ we will allow the  $\Phi_I $ to take any constant value at asymptotic infinity.

%%%%%%%%%%%%%%%%%%% 
\subsection{The supersymmetries}
%%%%%%%%%%%%%%%%%%% 

In five dimensions, supersymmetry generators come in symplectic pairs. While the $AdS$ metrics have Poincar\'e and conformal supersymmetries, the flow solutions will generically only preserve the  Poincar\'e supersymmetries.  In the  Poincar\'e patch these are precisely the supersymmetries that do not depend upon $t$ and $\vec x$ and so we will focus on those supersymmetries.    

On global $AdS$,  the supersymmetries generally have more complicated dependence on coordinates.    Consider the global $AdS_5$ metric:
\begin{equation}
ds_5^2 ~=~ R^2 \big[  \cosh^2 \lambda\,  dt^2 ~-~ \coeff{1}{4}\, \sinh^2 \lambda\,  (\sigma_1^2 + \sigma_2^2 + \sigma_3^2)  \, \ ~-~  d\lambda^2 \big]  \,,
  \label{AdSmet}
\end{equation} 
which corresponds to taking (\ref{fivemet2}) with:
\begin{equation}
r ~=~ a R  \log(\sinh \lambda)  \,, \qquad f ~=~ a \, \coth \lambda \,, \qquad   A~=~  \log(\coeff{R}{a} \sinh \lambda)    \,.
  \label{specialfns}
\end{equation} 
The symplectic pair of supersymmetries, $\hat{\epsilon}_{1}$ and $\hat{\epsilon}_{2}$,  in $AdS_5$ must satisfy:
\begin{equation}
\nabla_ \rho  \left(\begin{array}{c} 
 \hat \epsilon_1 \\ \hat \epsilon_2 
\end{array}\right)  ~+~ \left(\begin{array}{cc} 
0  & {1 \over 2 R}   \\ - {1 \over 2 R}  & 0
\end{array}\right)   \gamma_\rho \left(\begin{array}{c} 
\hat \epsilon_1 \\ \hat \epsilon_2 
\end{array}\right)  ~=~ 0~.
  \label{rndsusymat}
\end{equation} 
This is identically satisfied if we take the $\hat{\epsilon}_j$ to be independent of the $S^3$ and require:
\begin{equation}
\partial_t  \, \hat \epsilon_1 ~=~ -\coeff{1}{2} \, \hat \epsilon_2 \,, \qquad \partial_t  \, \hat \epsilon_2 ~=~ + \coeff{1}{2}\, \hat \epsilon_1  \,.
  \label{eptdep}
\end{equation} 

At infinity, the solution will become $AdS_5$ and we are going to allow solutions in which the gauge potentials, $\Phi_I$,  go to non-zero constants at infinity. Moreover a combination of the $U(1)$ gauge fields will represent the $U(1)$ $\mathcal{R}$-symmetry and so we need to incorporate the corresponding minimal couplings in the  $AdS_5$ background.  This can be done by inserting an arbitrary constant, $c$, into (\ref{eptdep}) and hence
we will seek symplectic pairs of supersymmetries that are independent of the homogeneous spatial three-surfaces and that satisfy
\begin{equation}
\partial_t  \, \hat \epsilon_1 ~=~ -\ds\frac{c }{a} \, \hat \epsilon_2 \,, \qquad \partial_t  \, \hat \epsilon_2 ~=~ + \ds\frac{c }{a} \,  \hat \epsilon_1  \,.
  \label{eptdepgen}
\end{equation} 
%

%%%%%%%%%%%%%%%%%%% 
\subsection{The supersymmetry conditions}
%%%%%%%%%%%%%%%%%%% 

To find five-dimensional supersymmetric bosonic backgrounds one sets the  variations of the spin-$1/2$ and spin-$3/2$ fields to zero.
From \cite{Gunaydin:1985cu}, the gravitational and scalar parts of these variations are:
\begin{eqnarray}
\delta\psi_{\mu a} &=&{\cal D}_\mu \epsilon_a ~-~ \coeff{1}{6}\, g\, W_{ab} \gamma_\mu \epsilon^b  ~-~ \coeff{1}{6}\, H_{\nu \rho\, ab} \big(\gamma^{\nu \rho} \gamma_\mu + 2 \gamma^\nu \delta^\rho_\mu  \big)\, \epsilon^b   \,,  \label{gravvar0}\\[8pt]
\delta\chi_{abc} &=&  \sqrt{2}~\Big[\gamma^\mu P_{\mu\, abcd} \, \epsilon^d ~-~ \coeff{1}{2} \, g\, A_{dabc}\, \epsilon^d  ~-~ \coeff{3}{4}\, \gamma^{\mu \nu}  H_{\mu \nu\, [ab}  \, \epsilon_{c]|}  \Big]  \,.  \label{gaugvar0}
\end{eqnarray} 

As in \cite{Freedman:1999gp}, the supersymmetries and superpotential can be isolated by looking at the eigenvalues of the $W_{ab}$ tensor.  Indeed there is a symplectic pair of such spinors with:
\begin{equation}
\begin{split}
W_{ab} \, \eta^b_{(k)} &~=~W\,  \eta^a_{(k)} \,, \qquad k = 1,2 \,,\\[8pt] 
\Omega_{ab} \,  \eta^b_{(1)}& =  -\eta^a_{(2)}  \,, \qquad   \Omega_{ab} \,  \eta^b_{(2)} = \eta^a_{(1)}   \,, 
\label{evecs}
\end{split}
\end{equation} 
where $\Omega$ is the symplectic form and $W$ is given by   (\ref{Wdefn}).
The complete supersymmetry of this system is then given by defining:
\begin{equation}
\epsilon^a ~\equiv~  \eta^a_{(1)} \,  \hat \epsilon_1 ~+~   \eta^a_{(2)} \,  \hat \epsilon_2    \quad  \Longrightarrow \quad \epsilon_a ~\equiv~  \Omega_{ab} \,\epsilon^b   =  - \eta^a_{(2)} \,  \hat \epsilon_1 ~+~   \eta^a_{(1)} \,  \hat \epsilon_2      \,, 
\label{susies}
\end{equation} 
where $\hat \epsilon_1$ and $\hat \epsilon_2$ are a symplectic pair of five-dimensional spinors on the space-time satisfying \eqref{eptdepgen}.

To write down the supersymmetry conditions, it is useful to define 
\begin{eqnarray}
\Lambda(n)   &\equiv& e^{-A} \bigg(\rho^2 \nu^{-2}   \Phi_1' + \rho^2  \nu^{2} \Phi_2'  +  \rho^{-4} \Phi_3'  +{n \over a}  \bigg)     \,,  
  \label{Lambdadefn} \\
\widetilde \Lambda  &\equiv&   -{g \over 2 f} \,e^{-A}  \, \Big[\,  {4 c \over  a g}~+~ ( \Phi_1+\Phi_2 -\Phi_3 ) \,
\cosh(2\, \varphi_1) ~+~ ( \Phi_1- \Phi_2+ \Phi_3 ) \,  \cosh(2\, \varphi_2) \nonumber \\ 
&& \qquad \quad ~+~ ( -\Phi_1+\Phi_2 +\Phi_3 ) \,  
\cosh(2\, \varphi_3) ~+~ ( \Phi_1+\Phi_2 +\Phi_3 ) \,\cosh(2\, \varphi_4) \,\Big]     \,,
  \label{Lambdatilde} 
\end{eqnarray} 
and
\begin{eqnarray}
X_\alpha  &\equiv&   {1\over 6} e^{-A}  \big(\rho^2 \nu^{-2}   \Phi_1' + \rho^2  \nu^{2} \Phi_2'  -2  \rho^{-4} \Phi_3'  \big)     \,,   \\[7pt]
 X_\beta &\equiv&   - {1\over 2} e^{-A} \big(\rho^2 \nu^{-2}   \Phi_1' -  \rho^2  \nu^{2} \Phi_2'    \big)        \,,
  \label{Xs} \\[7pt]
   \widetilde X^{(1)}     &\!\!\! \equiv\!\! \!& {g \over2 f} \,e^{-A}  ( \Phi_1 +  \Phi_2 -  \Phi_3 )  \sinh 2\varphi_1   \,,   \\  [7pt] 
   \widetilde X^{(2)}     &\!\!\! \equiv\!\! \!&  {g \over2 f} \,e^{-A}  ( \Phi_1 -  \Phi_2 +  \Phi_3 )  \sinh 2\varphi_2 \,,	   \\[7pt] 
 \widetilde X^{(3)}     &\!\! \!\equiv\! \! \!& {g \over2 f} \,e^{-A}  ( -\Phi_1 +  \Phi_2 +  \Phi_3 )  \sinh 2\varphi_3    \,,    \\ [7pt] 
 \widetilde X^{(4)}      &\!\!\! \equiv\!\! \!&  {g \over2 f} \,e^{-A}  ( \Phi_1 +  \Phi_2 +  \Phi_3 )  \sinh 2\varphi_4 \,.
  \label{Xtildes} 
\end{eqnarray} 

Assuming that the supersymetries only depend upon $t$  and $r$, with the $t$-dependence given by (\ref{eptdepgen}), the gravitino variations in the $t, r$ and other spatial directions give, respectively:
\begin{equation}
  \coeff {1}{ 2} (f' + f A')  \gamma^0 \gamma^4  \epsilon_a  ~-~    \coeff {1}{ 2}  \widetilde \Lambda \epsilon^a ~-~
 \coeff {1}{ 6} \, g\, W \gamma^0  \epsilon^a ~-~  \coeff {1}{ 3} \Lambda(0) \gamma^4  \epsilon_a ~=~ 0  \,,
\label{gvarform1}
\end{equation} 
\begin{equation}
f\,  \partial_r \epsilon_a  ~+~ \coeff {1}{ 6} \, g\, W \gamma^4  \epsilon^a ~+~  \coeff {1}{ 3} \Lambda(0) \gamma^0  \epsilon_a ~=~ 0  \,,
\label{gvarform2}
\end{equation} 
\begin{equation}
 \coeff {1}{ 2} \,  f A'  \gamma^0 \gamma^4  \epsilon_a    ~-~ \coeff {1}{ 6}\, g\, W \gamma^0  \epsilon^a ~+~  \coeff {1}{6} \Lambda(3)  \gamma^4  \epsilon_a ~=~ 0  \,.
\label{gvarform3}
\end{equation} 
Note that $-{1\over 2}  \widetilde \Lambda$ encodes the minimal couplings of the gauge fields to the supersymmetry and hence includes the parameter, $c$, that fixes the
$t$-dependence of the supersymmetries.

Subtracting (\ref{gvarform3}) from  (\ref{gvarform1})  gives 
\begin{equation}
  \coeff {1}{ 2}  f'    \gamma^0 \gamma^4  \epsilon_a  ~-~    \coeff {1}{ 2}  \widetilde \Lambda \epsilon^a   ~-~  \coeff {1}{ 2} \Lambda(1) \gamma^4  \epsilon_a ~=~ 0  \,.
\label{gvarform4}
\end{equation} 

The gaugino variations, $\delta \chi_{abc}$ break into six independent tensor components, corresponding to the six independent scalar fields.  These equations then reduce to:
\begin{eqnarray}
f \, \alpha'  \, \epsilon^a  ~-~  X_\alpha \gamma^0  \epsilon^a  ~+~    \coeff {1}{12} g\, {\partial W \over \partial \alpha} \, \gamma^4\, \epsilon^a  &=&  0  \,,
\label{gaugevar1} \\
f \, \beta'  \, \epsilon^a  ~-~  X_ \beta \gamma^0  \epsilon^a  ~+~    \coeff {1}{4} \, g\, {\partial W \over \partial \beta} \, \gamma^4\, \epsilon^a  &=& 0  \,,
\label{gaugevar2} \\ 
f \, \varphi_j'  \,\gamma^4\,  \epsilon^a  ~+~  \widetilde X_j \gamma^0  \epsilon_a  ~-~    \coeff {1}{2}\, g\, {\partial W \over \partial \varphi_j} \, \epsilon_a  &=&  0\,, \qquad j=1,\dots,4 \,.
\label{gaugevar3}
\end{eqnarray} 

In writing these equations we have explicitly assumed that the phases, $\theta_j$, are all constant.

%%%%%%%%%%%%%%%%%%% 
\subsection{Solving the supersymmetry variations}
%%%%%%%%%%%%%%%%%%% 

We now have to solve (\ref{gvarform2})--(\ref{gaugevar3}).  For the present we ignore the radial equation,  (\ref{gvarform2}), and solve the remaining equations, (\ref{gvarform3})--(\ref{gaugevar3}).   For the flows without electrostatic fields,  \cite{Freedman:1999gp, Khavaev:2000gb}, one is immediately led to a projection condition that gives $\hat \epsilon_2 =  \pm\gamma^4 \hat \epsilon_1$.  For the more general class of flows considered here one must allow for a ``dielectric projection'' condition on the spinors  \cite{Pope:2003jp, Gowdigere:2003jf, Pilch:2003jg}:
\begin{equation}
\left(\begin{array}{c} 
\hat \epsilon_1 \\ \hat \epsilon_2 
\end{array}\right)  ~+~ \left(\begin{array}{cc} 
\cos \xi \,  \gamma^0 & -\sin \xi \, \gamma^4 \\ \sin \xi \, \gamma^4 & \cos \xi \, \gamma^0
\end{array}\right)    \left(\begin{array}{c} 
\hat \epsilon_1 \\ \hat \epsilon_2 
\end{array}\right)  ~=~ 0 \,,
  \label{susymat1}
\end{equation} 
where $\xi = \xi(r)$.  With this projector one can then recast equations (\ref{gvarform3})--(\ref{gaugevar3}) as:
\begin{eqnarray}
\cos \xi &=&  {f' \over \Lambda(1)} ~=~ - {1\over 3}   {\Lambda(3) \over f \, A'}  ~=~ -  {2 \widetilde X^{(j)}  \over g \, (\partial_{\varphi_j} W)  }~=~ -  { X_ \alpha \over  f\, \alpha' } ~=~ -  { X_ \beta \over  f\, \beta' }   \,,
  \label{coseqn1} \\
\sin \xi &=& - {\widetilde \Lambda  \over \Lambda(1)} ~=~   {1\over 3}   {g\, W \over f \, A'} ~=~ -  {2f \varphi_j'  \over g \,(\partial_{\varphi_j} W) } ~=~ -  {g\over 12}   { (\partial_\alpha W) \over  f\, \alpha' }  ~=~ -  {g\over 4}   { (\partial_\beta W) \over  f\, \beta'}  \,.
  \label{sineqn1} 
\end{eqnarray} 
These represent  sixteen equations for the twelve arbitrary functions: $\alpha, \beta, \varphi_j$, $A, f, \Phi_I$ and $\xi$ but, as we will see, there is just the right degree of redundancy.

First observe that we have
\begin{equation}
{  \widetilde X^{(j)}  \over  \,(\partial_{\varphi_j} W) }  ~=~ {  \widetilde X^{(k)}  \over  \,(\partial_{\varphi_k} W) }  \,, 
  \label{algconstraints}
\end{equation} 
for all $j,k$.   For a generic flow in which all the $\varphi_j \ne 0$, (\ref{algconstraints}) are equivalent to:
\begin{equation}
\Phi_1  ~=~   \nu^4\, \Phi_2\,, \qquad  \Phi_3  ~=~  \rho^6\nu^2\, \Phi_2  \,.
  \label{constraint1}
\end{equation} 
For more specialized flows in which some of the $\varphi_j$'s vanish, or are equal, there may be additional possibilities, but we will ignore these in this paper.

Using (\ref{constraint1}), all six equations: 
\begin{equation}
\cos \xi ~=~   -  {2 \widetilde X^{(j)}  \over g \, (\partial_{\varphi_j} W)  }~=~ -  { X_ \alpha \over  f\, \alpha' } ~=~ -  { X_ \beta \over  f\, \beta' }   \,,
\end{equation} 
reduce to
\begin{equation}
\cos \xi ~=~   2\, f^{-1} \rho^2 \nu^2 \,e^{-A}  \, \Phi_2     \,,
  \label{coseqn2}
\end{equation} 
There remain eleven equations for the ten unknown functions: $\alpha, \beta, \varphi_j$, $A, f, \Phi_2$ and $\xi$.
 
Using  (\ref{constraint1}) one also finds:  
\begin{equation}
\begin{split}
 \Lambda(n)  &~=~  3\,  e^{-A}\Big[ {d \over dr} \big(\rho^2 \nu^2 \, \,  \Phi_2 \big)~+~{ n\over 3 a}  \Big]  \,, \\[8pt] 
 \widetilde \Lambda  &~=~     -{2 c  \over   a f} \,e^{-A} ~+~  {2\,g \over   f} \,e^{-A}  \, \rho^2 \nu^2 \, W\,  \Phi_2     \,.
  \label{Lambdared1}
\end{split}
\end{equation} 
Using this in the second identity for $\cos \xi$ in   (\ref{coseqn1})  yields the differential equation:
\begin{equation}
 {d \over dr} \big(\rho^2\nu^2\, e^{2 A} \, \Phi_2 \big) ~=~  - {1 \over  a} \, e^{2 A}  \,.
  \label{identb}
\end{equation} 

Now use the following identity from the expressions for $\sin \xi$:
\begin{equation}
{\widetilde \Lambda  \over \Lambda(1) } ~=~  - {1\over 3}   {g\, W \over f \, A'}  \,, 
  \label{sinident2}
\end{equation} 
which is equivalent to:
\begin{equation}
A'  ~=~  - {g \over 3 c}  \, W  \,.
  \label{Aflow1}
\end{equation} 
One then obtains
\begin{equation}
f \, \sin \xi ~=~  - c \,.
  \label{fsinident}
\end{equation} 
From the last three expressions for $\sin \xi$ one obtains the steepest descent equations:
\begin{equation}
\alpha'  ~=~    {g \over 12 c}  \, {\partial W \over \partial \alpha} \,, \qquad   \beta'  ~=~    {g \over 4 c}  \, {\partial W \over \partial \beta} \,,\qquad   \varphi_j'  ~=~    {g c \over 2  f^2}  \, {\partial W \over \partial \varphi_j}  \,.
\label{steep1}
\end{equation} 
Note the presence of the factor of $f^{-2}$  in the last of these equations.

Equations   (\ref{Aflow1}),  (\ref{fsinident}) and  (\ref{steep1}) now account for all the equations coming from the expression for  $\sin \xi$ while   (\ref{coseqn2}) and   (\ref{identb}) account for all but one of the equations for $\cos \xi$.  There are thus two remaining equations:   $\cos^2 \xi+ \sin^2 \xi = 1$ and 
\begin{equation}
f' =~    \Lambda(1) \, \cos\xi  \,.
  \label{fdiif1}
\end{equation} 
Using   (\ref{coseqn2}) and   (\ref{identb}) this can be written:
\begin{equation}
 {d \over dr} \Big[ f^2 ~-~ 4 e^{-2A}\big( \rho^2\nu^2\,  \Phi_2 \big)^2  \Big] ~=~  {d \over dr} \Big[ f^2 ~-~ f^2\cos^2 \xi   \Big] ~=~  0 \,, 
  \label{fdiif2}
\end{equation} 
which is precisely consistent with   (\ref{fsinident}).  Finally, note that $\cos^2 \xi+ \sin^2 \xi = 1$  is equivalent to:  
\begin{equation}
 \rho^2\nu^2\, e^{2A} \Phi_2   ~=~ - \coeff{1}{2}\, e^{3A}  \sqrt{f^2 - c^2}    \,, 
  \label{csqssq}
\end{equation} 
and hence we can write  (\ref{identb}) as an equation for the evolution of $f$:
\begin{equation}
 {d \over dr} \Big[ \, e^{3A}  \sqrt{f^2 - c^2}  \,  \Big] ~=~ + {2 \over a} \,e^{2A} \,.
  \label{fdiif3}
\end{equation} 
Note that  the sign choice in  (\ref{csqssq}), and hence in (\ref{fdiif3}), has been made  so that $f$ behaves in the appropriate manner in the UV limit, $r \to + \infty$, to obtain an $AdS$ space as in \eqref{AdSflowcoords}.

%%%%%%%%%%%%%%%%%%% 
\subsection{The radial dependence of the supersymmetries}
%%%%%%%%%%%%%%%%%%% 

It simplifies the radial equation, (\ref{gvarform2}), if one multiplies  (\ref{gvarform1}) by $\gamma^0 \gamma^4$ and subtracts it from  (\ref{gvarform2})   to obtain:
\begin{equation}
f\,  \partial_r \epsilon_a  ~+~  \coeff {1}{ 2} (f' + f A')   \epsilon_a    ~+~    \coeff {1}{ 2}  \widetilde \Lambda  \gamma^0 \gamma^4 \epsilon^a  ~=~ 0  \,.
\label{gvarform6}
\end{equation} 

Define $\hat{\epsilon}_j^{(0)}$ by
\begin{equation}
  \left(\begin{array}{c} 
\hat \epsilon_1 \\ \hat \epsilon_2 
\end{array}\right)   ~=~ f^{1 \over 2}  \, e^{{A \over 2} } \, \left(\begin{array}{cc} 
\cos\frac{\xi}{2}  & \sin\frac{\xi}{2} \, \gamma^0 \gamma^4 \\ -\sin\frac{\xi}{2} \,  \gamma^0 \gamma^4 & \cos \frac{\xi}{2} 
\end{array}\right)    \left(\begin{array}{c} 
\hat \epsilon_1^{(0)} \\ \hat \epsilon_2^{(0)}  
\end{array}\right)   \,.
  \label{susymat2}
\end{equation} 
Then the projection condition, (\ref{susymat1}), and radial equation, (\ref{gvarform6}), reduce to the elementary form:
\begin{equation}
\big( \oneone +  \gamma^0 \big) \, \hat\epsilon_j^{(0)}~=~ 0 \,, \qquad  \partial_r \, \hat\epsilon_j^{(0)}~=~ 0 \,, \qquad j =1,2 \,.
\label{susysimple1}
\end{equation} 
The differential equation for the $\hat\epsilon_j^{(0)}$ follows by using (\ref{coseqn1})  and (\ref{sineqn1}) to set $\widetilde \Lambda = - f' \tan \xi$  in   (\ref{gvarform6}) and then using \eqref{fsinident}.

%%%%%%%%%%%%%%%%%%% 
\subsection{The  flow equations}
%%%%%%%%%%%%%%%%%%% 

Pulling the results together, the flows are given by solving the first order system:
\begin{equation}
\alpha'  ~=~    {g \over 12 c}  \, {\partial W \over \partial \alpha} \,, \qquad   \beta'  ~=~    {g \over 4 c}  \, {\partial W \over \partial \beta} \,,\qquad   \varphi_j'  ~=~    {g c \over 2  f^2}  \, {\partial W \over \partial \varphi_j}  \,.
\label{steep}
\end{equation} 
\begin{equation}
A'  ~=~  - {g \over 3 c}  \, W \,, \qquad  {d \over dr} \Big[ \, e^{3A}  \sqrt{f^2 - c^2}  \,  \Big]  ~=~   {2 \over a} \,e^{2A} \,,
  \label{metflow}
\end{equation} 
to obtain $\alpha, \beta, \varphi_j$, $A$, and $f$, where $c$ is a constant.  

The electrostatic potentials are then given by:
\begin{equation}
\Phi_1   =  -\coeff{1}{2}\, e^{A}  \rho^{-2}\nu^{2} \sqrt{f^2 - c^2}    \,,  \quad \Phi_2   = - \coeff{1}{2}\, e^{A}  \rho^{-2}\nu^{-2} \sqrt{f^2 - c^2}    \,,  \quad \Phi_3   = - \coeff{1}{2}\, e^{A}  \rho^{4}   \sqrt{f^2 - c^2}    \,,  
  \label{Phians1}
\end{equation} 
and the polarization angle is given by: 
\begin{equation}
\sin \xi   ~=~ - c f^{-1}     \,.
  \label{xians1}
\end{equation} 

These flow equations have many solutions with diverging scalar fields. Some of them correspond to flows driven by relevant perturbations in the dual field theory and some are driven by condensates. While many of these flows may represent physically interesting solutions in ten dimensions, or in the dual field theory, we will now focus on flows between the supersymmetric fixed points.

%%%%%%%%%%%%%%%%%%%%%%%%%%%%%%%%%%%%%%%%%%%%%%%%%%%%%%%%%  
\section{Charged clouds}
%%%%%%%%%%%%%%%%%%%%%%%%%%%%%%%%%%%%%%%%%%%%%%%%%%%%%%%%%  

%%%%%%%%%%%%%%%%%%% 
%\subsection{UV and IR asymptotics}
%%%%%%%%%%%%%%%%%%%  

To get the flow to $AdS_5$ at infinity one  must have $A \sim {r \over L}$ and so one must take $f \to c$ and hence $\xi \to -{\pi \over 2}$ as $r \to \infty$.  The supersymmetry projector 
(\ref{susymat1}) reduces to precisely that of \cite{Freedman:1999gp}, which, given (\ref{gammconv2}), is the standard $D3$ brane projection. 
To get the canonically normalized time and radial coordinate on $AdS_5$ one takes $c=1$.

If one takes the UV limit to be the maximally supersymmetric critical point with $\rho = \nu =1, \varphi_j =0$ and $W= -{3 \over 2}$, then  we have:
\begin{equation}
c ~=~ 1 \,,  \qquad g ~=~ {2 \over L} \,,  \qquad  \Phi_1,  \Phi_2 ,  \Phi_3   ~\to~  - {L \over 2  a}  \,, \qquad  f ~\sim~ 1 ~+~ \ds\frac{ L^2}{2 a^2} \, e^{-{2r \over L}}  \,. 
  \label{UVasymp1}
\end{equation} 
The flow thus starts out in exactly the same manner as those of
\cite{Freedman:1999gp,Khavaev:2000gb,Freedman:1999gk,Pilch:2000ue,Warner:1999kh}. 

It may, at first seem surprising that supersymmetry requires non-zero electrostatic potentials at infinity.  Coulomb potentials are dual, in the holographic field theory, to chemical potentials and constant values represent negative mass terms in the field theory.    Such terms would normally destabilize the vacuum and cause the formation of a condensate, but  
the holographic dual field theory for {\it global} $AdS$ is a conformal field theory on $S^3 \times \IR$ and thus all scalars in that field theory must be conformally coupled (see, for example, \cite{Yamada:2006rx}) with mass terms of  the form ${1 \over 6} R \phi^2 =  {1 \over  a^2} \phi^2$ on $S^3$. To get a supersymmetric flow, the negative mass term from the chemical potential must cancel the positive mass from the scalar curvature of $S^3$ and hence one should expect the Coulomb fields to behave exactly as in (\ref{UVasymp1}).

The flow to the non-trivial IR fixed point has some somewhat different features from those of \cite{Freedman:1999gp}.  The crucial difference is that $f$ diverges in the infra-red and thus greatly slows the flow of $\varphi_j$.

The flow of interest has $\chi \equiv \varphi_1 \ne 0$,   $\beta = \varphi_j =0$, $j \ge 2$ and $\Phi_1 = \Phi_2$.  In terms of the charged scalar fields, $\zeta_j$, one only has $\zeta_1 \ne 0$. The IR fixed point is then characterized by:
\begin{equation}
\begin{split}
\alpha &~=~ \alpha_* \equiv \coeff{1}{6} \, \log(2)  \,, \qquad \chi ~=~ \chi_* \equiv  \coeff{1}{2} \, \log(3) \,, \\[8pt] 
W   &= -2^{2/3} \ \ \Rightarrow \ \  L_* =   \coeff{3}{2} \,  2^{-2/3}   L\,,  \qquad e^{A} ~\sim~e^{r/L_*}  \,, 
\label{IRFP}
\end{split}
\end{equation} 
with $r \to - \infty$.  

For regular solutions as $r \to - \infty$ one must exclude all the homogeneous solutions in the equations for $\Phi_j$ and $f$ ((\ref{identb}) and (\ref{steep})) and thus one has 
\begin{equation}
e^{A} ~\sim~e^{+ { r \over L_*}}    \,, \qquad  f ~\sim~ \coeff{L_*}{a}  \, e^{-{ r \over L_*}}  \,,  \qquad  \Phi_1=  \Phi_2 ~\to~  - 2^{-4/3}\,  \coeff{L_*}{a}\,, \quad  \Phi_3   ~\to~  -2^{-1/3}\,  \coeff{L_*}{a}   \,. 
  \label{IRasymp1}
\end{equation} 
as  $r \to - \infty $.  Note that in this limit the $U(1)\times U(1)$ gauge group of the domain wall solution is spontaneously broken to $U(1)$, $\Phi_3 =  2\Phi_1 =2 \Phi_2$, and from (\ref{Lagrangian})  one sees that the minimal coupling to $\zeta_1$ vanishes for these values  of the potential.  Thus the flow has completely expelled the charge carried by  $\zeta_1$.

To understand the infra-red flow in more detail, observe that the superpotential has the expansion about the point (\ref{IRFP}) given by:
\begin{equation}
W ~=~  -2^{2/3}  \big[ 1 ~+~ 4\, (\alpha  - \alpha_*)^2 -  4\, (\alpha  - \alpha_*)( \chi  - \chi_*)   ~+~ \dots \big]  \,, 
\label{IRWexp}
\end{equation} 
where $ + \dots $ indicates terms of cubic order.  This critical point is a saddle point with one positive eigenvalue and one negative eigenvalue.  However, because of the presence of the $f^{-2}$ in (\ref{steep}) the asymptotic behavior of the flow is not so simple.  For large, negative $r$ one has 
\begin{equation}
{d \alpha \over dr}   ~=~   -{1 \over  L_* }\,   \big[ 2\, (\alpha  - \alpha_*) -  ( \chi  - \chi_*)   ~+~ \dots \big]  \,,  \qquad 
{d \chi \over dr}    ~=~    {6 a^2\over   L_*^3 }\,  e^{ {2 r \over L_*}}\,  (\alpha  - \alpha_*)  ~+~ \dots  \,, 
\label{IRdiffeqns}
\end{equation} 
where $ + \dots $ indicates terms of quadratic order.  Note the exponential damping, coming from $f^{-2}$, in the second equation.

There is one obvious solution in the neighborhood of the critical point, and it has 
\begin{equation}
(\alpha  - \alpha_*) ~\sim~ k_0 \, e^{ - {2 r \over L_*}}  \,,  \qquad  ( \chi  - \chi_*) ~ \sim ~{6 a^2\over   L_*^3 }\,k_0 \, r    \,,
\label{IRflow1}
\end{equation} 
for some constant, $k_0$. However, this solution diverges as $r \to - \infty$ and corresponds to a flow away from the critical point.

The flow that approaches the critical point requires some delicate fine tuning of the path so that $ ( \chi  - \chi_*) \sim 2 (\alpha  - \alpha_*) $ along the approach.  Indeed, suppose one starts from an extremely small displacement away from this critical point, then one has  
\begin{equation}
(\alpha  - \alpha_*) ~=~ c_0 \big(1 ~+~ \coeff{3 a^2}{4 L_*^2}  \, e^{  {2 r \over L_*}}~+~ \dots  \big)  \,,  \qquad  ( \chi  - \chi_*)  ~=~ 2 \, c_0 \big(1 ~+~  \coeff{3 a^2}{2 L_*^2}  \,  e^{  {2 r \over L_*}} ~+~ \dots \big)      \,,
\label{IRflow2}
\end{equation} 
where $c_0$ is an arbitrarily small constant and $ + \dots $ indicates terms that vanish faster than  $e^{  {2 r \over L_*}}$ as $r \to - \infty$.    Observe that this solution relaxes, exponentially slowly to $(\alpha  - \alpha_*)  = {1 \over 2} ( \chi  - \chi_*)   = c_0$.  It thus requires the higher order, non-linear terms in the flow equations to drive the flow onto the  fixed point.  Thus the flow approaches the fixed point extremely slowly, probably more slowly than any simple exponential decay.

\begin{figure}[!ht]
\begin{center}
\includegraphics[width=8.5cm]{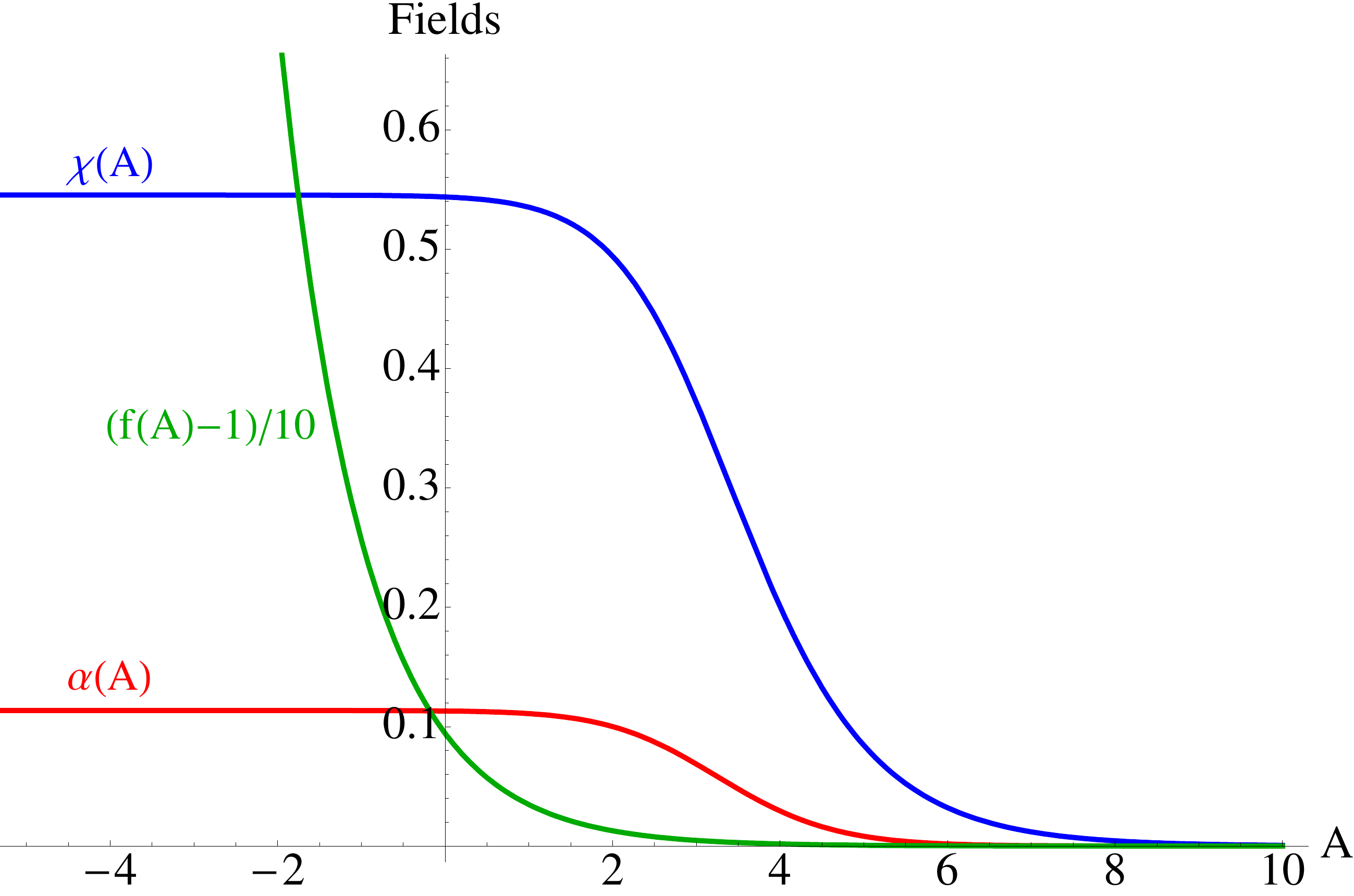}
\caption{{\it The profiles of the scalar functions $\alpha(A)$, $\chi(A)$ obtained by solving equation (\ref{steep}) and the metric function $f(A)$ obtained by solving equation (\ref{metflow}). For clarity, we have plotted the functions $\alpha(A)$, $\chi(A)$ and $(f(A)-1)/10$.}}
\label{fig1}
\end{center}
\end{figure}

We have solved for this flow numerically, and we have shown the profiles of the various fields in Fig.~\ref{fig1}. Note that the variations of these fields are confined within a relatively small region of the $A$-space. These variations occur largely before $A(r)$ passes through zero, in other words, before the point where the holographic scale becomes of the order of the scale of the $S^3$ ``box'' in which the field theory lives.

%%%%%%%%%%%%%%%%%%%%%%%%%%%%%%%%%%%%%
%%%%%%%%%%%%%%%%%%%%%%%%%%%%%%%%%%%%%

\begin{figure}[!ht]
\begin{center}
\subfigure[] {\includegraphics[angle=0,
width=0.45\textwidth]{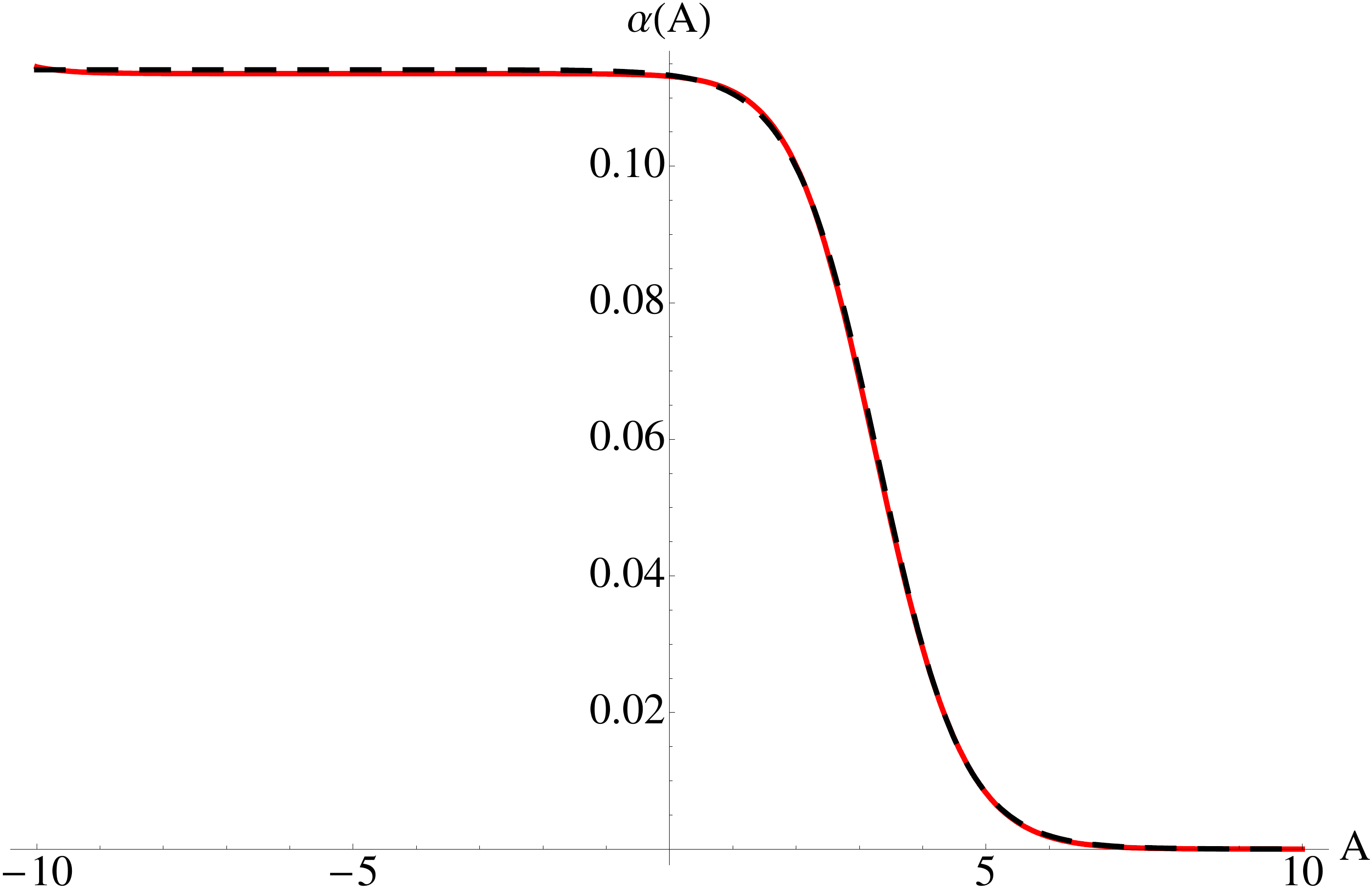} \label{fig:alpha}}
\subfigure[] {\includegraphics[angle=0,
width=0.45\textwidth]{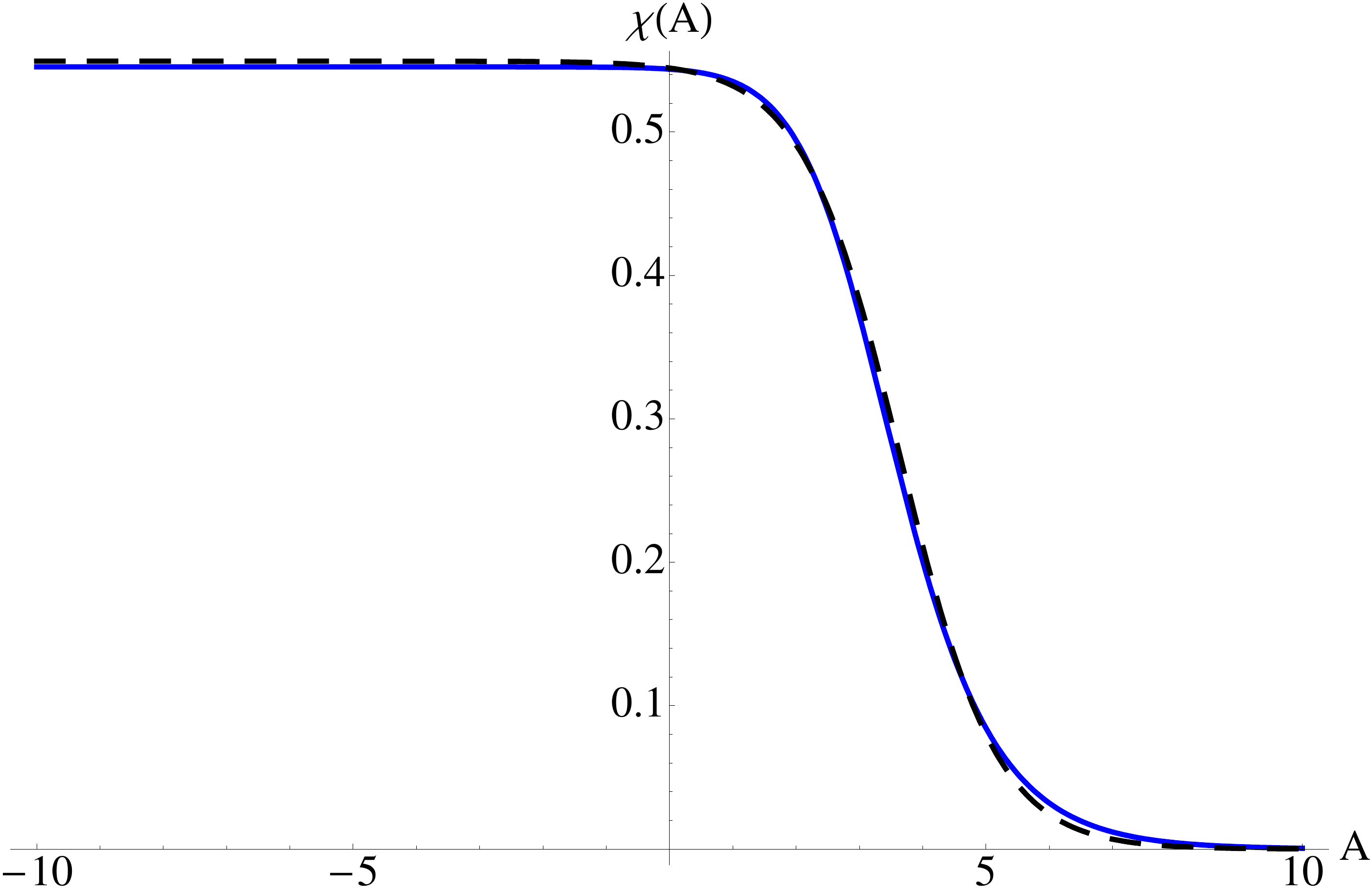} \label{fig:chi}}
\subfigure[]{\includegraphics[angle=0,
width=0.45\textwidth]{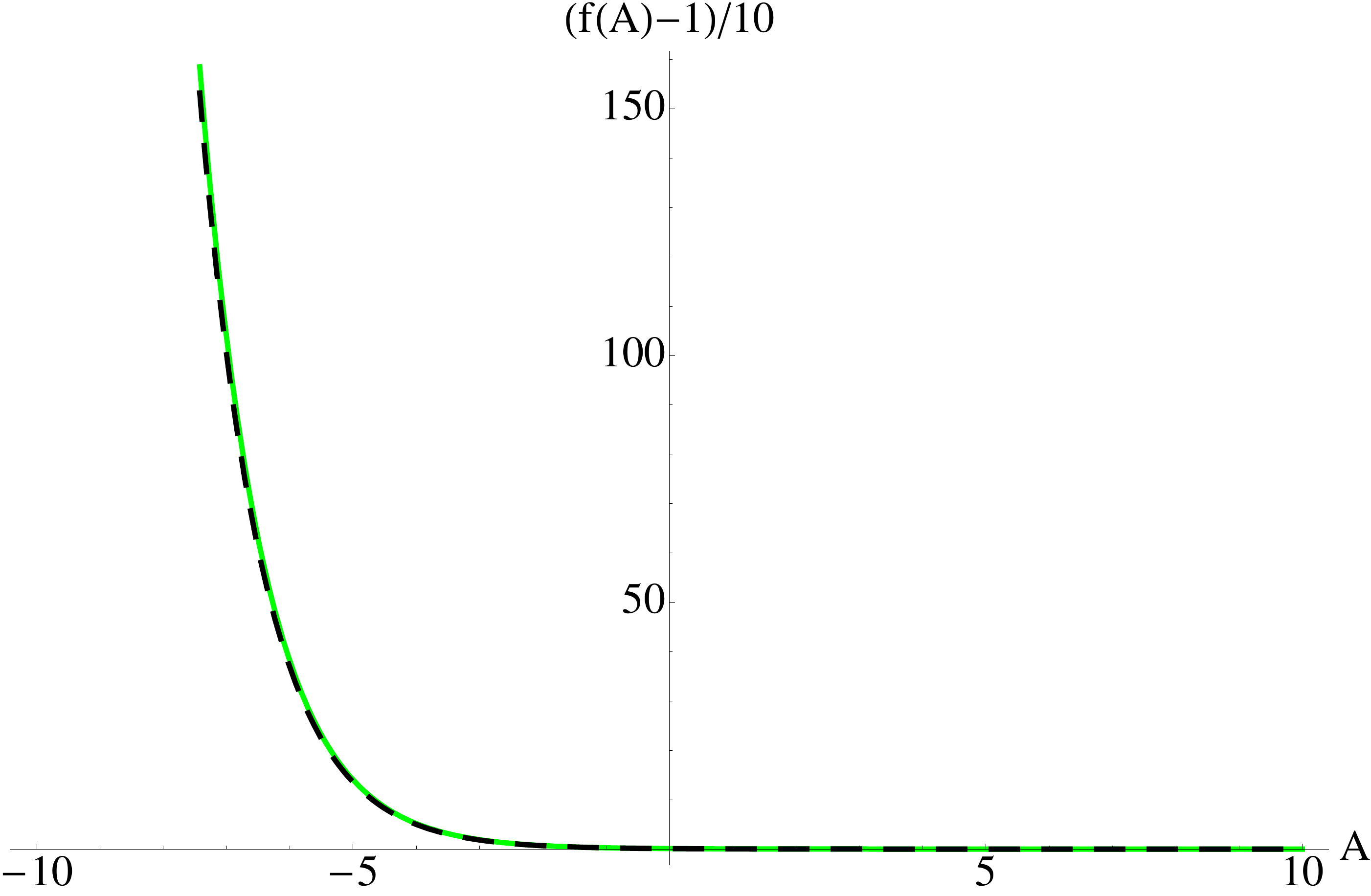} \label{fig:f}}
\caption{{\it The black dashed lines are the functions given in equation (\ref{exactalpha})--(\ref{exactf}). Clearly they resemble the numerical solutions very closely.}}
\label{exacts}
\end{center}
\end{figure}

We have not been able to solve equations (\ref{steep}) and (\ref{metflow}) analytically. However, from the numerical profiles we are able to guess the following analytical expressions which fit the numerical solutions rather well:
\begin{eqnarray}
\alpha(A) & = & - \frac{\alpha_{*}}{2} \left[ \tanh\left( 0.75 \left( A - 3.3 \right)\right) - 1 \right] \ , \label{exactalpha} \\
\chi(A) & = & - \frac{\chi_{*}}{2} \left[ \tanh\left( 0.65 \left( A - 3.65 \right)\right) - 1 \right] \ , \label{exactchi} \\
\frac{f(A) -1}{10} & = & \frac{1}{2} e^{- \left( A + 1.7 \right)} \ , \label{exactf}
\end{eqnarray}
where $\alpha_*$ and $\chi_*$ are the values of the scalars at the IR fixed point (given in equation (\ref{IRFP})). We have pictorially demonstrated this fit in Fig.~\ref{exacts}. We also find that the numerical constants appearing in equation (\ref{exactalpha})-(\ref{exactf}) are not fine-tuned, namely these fitting curves are not extremely sensitive to their respective values quoted above.

%%%%%%%%%%%%%%%%%%%%%%%%%%%%%%%%%%%%%
%%%%%%%%%%%%%%%%%%%%%%%%%%%%%%%%%%%%%

%%%%%%%%%%%%%%%%%%%%%%%%%%%%%%%%%%%%%%%%%%%%%%%%%%%%%%%%%  
\section{Charged black holes}  
%%%%%%%%%%%%%%%%%%%%%%%%%%%%%%%%%%%%%%%%%%%%%%%%%%%%%%%%%

In the previous sections we suppressed the homogeneous solutions  in the equations for $\Phi_J$ and $f$, (\ref{identb}) and (\ref{steep}), because the homogeneous solutions generically generate singularities.   Since we are dealing with highly symmetric, charged solutions, it should come as no surprise that these singular solutions look somewhat like black holes.   They are not actually black holes in that they have naked singularities.  Indeed, we have been focussing on solutions with four supersymmetries and the corresponding black-hole solutions in $AdS_5$ are known to be singular \cite{Behrndt:1998ns}, with the only known supersymmetric black hole with a regular horizon possessing only two supersymmetries  \cite{Gutowski:2004ez,Gutowski:2004yv}.

%%%%%%%%%%%%%%%%%%%%%%%%%%%%%%%%%%%%%%
\subsection{The singular ``black hole'' solutions}
\label{STUBPS}
%%%%%%%%%%%%%%%%%%%%%%%%%%%%%%%%%%%%%%

In five dimensions, the STU model is simply the $\Neql2$ supersymmetric theory consisting of a graviton multiplet coupled to two vector mutiplets and can be obtained from our formulation by setting all the hypermultiplet scalars to zero, $\varphi_j =0$.    We consider the global  $AdS_5$ metric, (\ref{fivemet2}), and  introduce a new coordinate, $\eta$, defined via the implicit coordinate transformation:
\begin{equation}
dr = \ds\frac{2 c}{g}~\ds\frac{ \eta}{ [(\eta^2+q_1)(\eta^2+q_2)(\eta^2+q_3)]^{1/3}} d\eta \,.
\end{equation}
One can then show that the following represents a solution to the BPS equations:
\begin{eqnarray}
e^{2A(\eta)} &=& \ds\frac{g}{2  c} [(\eta^2+q_1)(\eta^2+q_2)(\eta^2+q_3)]^{1/3} ~, \qquad\qquad \varphi_{j} = 0~,\\[8pt]
\alpha(\eta) &=& \ds\frac{1}{12} \log\left( \ds\frac{(\eta^2+q_1)(\eta^2+q_2)}{(\eta^2+q_3)^2} \right)~, \qquad\qquad \beta(\eta) = \ds\frac{1}{4} \log\left( \ds\frac{\eta^2+q_2}{\eta^2+q_1} \right)~,  \\[8pt]
\Phi_1 &=& - \ds\frac{c}{ag} \ds\frac{\eta^2}{\eta^2+q_1} ~, \qquad\qquad \Phi_2 = - \ds\frac{c}{ag} \ds\frac{\eta^2}{\eta^2+q_2}~,  \qquad\qquad \Phi_3 = - \ds\frac{c}{ag} \ds\frac{\eta^2}{\eta^2+q_3}~,\\[8pt]
f^{2}(\eta) &=& c^2 + \ds\frac{8 c^3}{g^3 a^2} ~\ds\frac{\eta^4}{(\eta^2+q_1)(\eta^2+q_2)(\eta^2+q_3)}~.
\end{eqnarray}
Define the functions
\begin{equation}
H_{i} (\eta) = \ds\frac{g}{2c} \left(1 + \ds\frac{q_i}{\eta^2}\right)~, \qquad\qquad F(\eta) = \ds\frac{1}{a^2} + c^2 \eta^2 H_1H_2H_3 \,, 
\end{equation}
and then one can rewrite the solution in the more canonical form: 
\begin{equation}
ds^2_5 = \ds\frac{F}{(H_1H_2H_3)^{2/3}} dt^2 -  \coeff{1}{4} a^ 2 (H_1H_2H_3)^{1/3} \eta^2 (\sigma_1^2 + \sigma_2^2 + \sigma_3^2)  - \ds\frac{(H_1H_2H_3)^{1/3}}{F} d\eta^2~,
\end{equation}
with gauge fields and scalars given by:
\begin{equation}
A_{(i)} = - \ds\frac{1}{2a} \ds\frac{1}{H_i} dt~,\qquad\qquad X^{(i)} =  \ds\frac{(H_1H_2H_3)^{1/3}}{H_i}~.
\end{equation}
These are precisely the BPS solutions of the STU model with three different charges discussed in \cite{Behrndt:1998ns, Behrndt:1998jd, Cvetic:1999ne, Myers:2001aq}. Note that, as required by the STU model, we have $X^{(1)}X^{(2)}X^{(3)}=1$, with 
\begin{equation}
X^{(1)} = \rho^{-2}\nu^2~, \qquad X^{(2)} = \rho^{-2}\nu^{-2}~, \qquad X^{(3)} = \rho^{4}~.
\end{equation}
The solution can be analytically continued through $\eta^2 = 0$ and has a naked singularity at $\eta^2  = - q$, where $q$ is the smallest of the three charges \cite{Myers:2001aq}. It is clear that if one takes the parameter $a$ to infinity one will get the corresponding solution in the Poincar\'e patch.

In terms of flows driven by the superpotential, $W$, this solution represents a ``flow to Hades'' \cite{Warner:1999kh} in that $|W| \to \infty$. It is also a rather simple such flow in that it does not involve the hypermultiplets since $\varphi_j=0$.    There are obviously large numbers of similar flows with $\varphi_j \ne 0$  and it would be interesting to understand them in terms of ``black-hole-like" solutions.  These will certainly include the generalized Coulomb branch flows \cite{Khavaev:2001yg} associated with the non-trivial fixed point theory.

The scalar potential for this solution is
\begin{equation}
\mathcal{P} = - g^2 \ds\frac{3\eta^2 + q_1+q_2+q_3}{ [(\eta^2+q_1)(\eta^2+q_2)(\eta^2+q_3)]^{1/3}}~.
\end{equation}
It is clear that $\mathcal{P} \to -\infty$ as one approaches the naked singularity but it is important to note that the scalar potential is bounded above throughout the whole solution. Following  \cite{Gubser:2000nd}, one can argue that such singular five-dimensional solutions are ``good" (i.e. physically relevant) and the singularity should be resolved by a distribution of D-branes once the solutions is uplifted to type IIB supergravity. Indeed there is a known physical interpretation of this solution as giant gravitons on $AdS_5\times S^5$ \cite{ Myers:2001aq}.

%%%%%%%%%%%%%%%%%%%%%%%%%%%%%%%%%%%%%%
\subsection{Solutions with constant scalars} 
\label{PWBHBPS}
%%%%%%%%%%%%%%%%%%%%%%%%%%%%%%%%%%%%%%

It is also interesting to see what classes of supersymmetric solutions can be generated from a critical point of the superpotential.  That is, suppose that all scalars take constant values $(\alpha_*,\beta_*,\varphi_j^{*})$ corresponding to an extremum of the superpotential, $W$, whose value at the extremum is given by $W_{*}$.  This automatically solves the BPS equations for $\alpha$, $\beta$ and $\varphi_j$ given in (\ref{steep}). It is then elementary to obtain the solution for the rest of the fields:
\begin{eqnarray}
A(r) & = & - \ds\frac{g}{3c} W_{*} ~  r + A_0~, \label{BPSnaked1}\\[8pt]
\Phi_2 (r) & = &  \ds\frac{ 3 c}{2 a g W_*} \ds\frac{1}{(\rho_* \nu_*)^2} + C_1 e^{-2A(r)} ~, \\[8pt]
\Phi_1 (r) &=& \nu_*^4 \Phi_2(r) ~, \qquad\qquad \Phi_3(r) = \rho_*^6 \nu_*^2 \Phi_2(r) ~,\\[8pt]
f^2 (r) & = & c^2 + 4 (\rho_* \nu_*)^4 e^{-2A(r)} (\Phi_2(r))^2 ~,
\label{BPSnaked4}
\end{eqnarray}
where $C_1$ and $A_0$ are integration constants. Note that we have included the homogeneous parts in solving equations (\ref{metflow}).

Since the three Coulomb potentials are the same, up to  overall scales, this means that we are dealing with the $STU$ black hole (naked singularity, to be more precise) with $q_1 = q_2 =q_3 = q$.  The fact that we have been forced into the minimal solution arises from the generic value of the $\varphi_j^*$ and the resulting constraints (\ref{constraint1}).  We thus have the rather simple metric:
\begin{equation}
ds^2_5 = \ds \widetilde{F} H^{-2} d\tau^2 -  \coeff{1}{4} a^ 2 H \eta^2 (\sigma_1^2 + \sigma_2^2 + \sigma_3^2) - \ds H\widetilde{F}^{-1} d\eta^2~,
\end{equation}
with 
\begin{eqnarray}
H(\eta) & = & \ds\frac{g}{2c} \left( 1 + \ds\frac{q}{\eta^2} \right) \ , \quad \widetilde{F}(\eta) = \ds\frac{1}{a^2} + \ds\frac{4 c^2  W_{*}^2 }{9} \eta^2 H^3(\eta)~, \quad e^{2A(r) }  =   \ds\frac{g}{2c} \left(\eta^2 + q \right)~,  \\
\tau &=& - \ds\frac{3}{2 W_{*}} t~, \qquad\qquad e^{2A_0} = \ds\frac{g}{2c}~, \qquad\qquad
C_1 =  - \ds\frac{3 q}{4 a W_{*}  \left( \rho_* \nu_* \right)^2 }   ~. 
\end{eqnarray}
The vector potentials are given by:
\begin{equation}
A_{(1)} = - \ds\frac{\nu_{*}^2}{2 a  \rho_*^2} ~H^{-1} d\tau~,\qquad A_{(2)} =  -\ds\frac{1}{2 a (\rho_* \nu_{*})^2}~ H^{-1} d\tau~, \qquad A_{(3)} = -\ds\frac{\rho_{*}^4}{2 a  } ~H^{-1} d\tau~,
\end{equation}
and the scalars have the constant values: 
\begin{equation}
X^{(1)} =  (\rho_*)^{-2}(\nu_{*})^2~,\qquad X^{(2)} =  (\rho_*)^{-2}(\nu_{*})^{-2}~, \qquad X^{(3)} =  (\rho_*)^{4}~. \label{constSTUscalars}
\end{equation}

The underlying $STU$ model has a scale invariance:  $A_{(j)} \to \mu_j A_{(j)}$, $X^{(j)} \to \mu_j  X^{(j)}$, provided that $\mu_1 \mu_2 \mu_3 =1$.  Taking $\mu_1 =   \rho_* ^2\nu_*^{-2}$,  $\mu_2 =   \rho_* ^2\nu_*^{2}$ and  $\mu_3 =   \rho_* ^{-4}$ will scale this solution to precisely the standard, minimal $STU$ black hole.   

This mathematical statement, however, misses a much more important physical point:  The  solution at the non-trivial critical (PW) point corresponding to (see \eqref{IRFP})
\begin{equation}
\alpha_* = \ds\frac{1}{6} \log 2~, \qquad \chi_*\equiv \varphi_1^* = \pm \ds\frac{1}{2} \log 3~, \qquad \beta_* = \varphi_2^*=\varphi_3^*=\varphi_4^*=0~, \qquad W_{*} = - 2^{2/3}~,
\label{critvals1}
\end{equation}
represents a new solution to the five-dimensional theory coupled to a charged hypermultiplet.   The fact that this solution collapses to the standard $STU$ form becomes apparent if one remembers that the (Abelian) gauge group at the PW point has 
\begin{equation}
A_{(1)} = A_{(2)}~,\qquad\qquad A_{(3)} = 2 A_{(1)}~. \label{pwacon} 
\end{equation}
Now recall that the only non-zero charged scalar, $\varphi_1$,   couples to $(A_{(1)} + A_{(2)} -A_{(3)})^2$ in (\ref{Lagrangian}) and thus $\varphi_1$ decouples from everything except the potential.  As a result, the only r\^ole played by the non-trivial hypermultiplet scalar in the solution at the critical point is to arrange that there is a critical point and to set the scale via the cosmological constant.    

There are thus two interesting charged solutions arising from the five-dimensional solution coupled to a charged hypermultiplet:  The more symmetric charged solution, at which all the charged scalars vanish, and the charged solution arising from the PW fixed point with a constant charged scalar $\varphi_1$.  In the next section we will exhibit a supersymmetric charged domain wall that interpolates between these two solutions. 

From a ten-dimensional perspective in IIB supergravity, the standard $STU$ black hole and the new class of solutions will also look extremely different.   As is evident from the earlier, uncharged solutions and flows  \cite{Pilch:2000ej, Pilch:2000fu}, the new solutions will involve non-trivial internal fluxes and further distortion of the metric on $S^5$.  From a ten-dimensional perspective, the addition of Coulomb potentials corresponds to adding angular momenta on the $S^5$ \cite{Cvetic:1999xp} and so the new family of charged black-hole-like solutions based on the PW critical point will lift to new families of giant gravitons in ten dimensions \cite{McGreevy:2000cw,Myers:2001aq}.

%%%%%%%%%%%%%%%%%
\subsection{Flows between BPS naked singularities}
%%%%%%%%%%%%%%%%%

We now consider the effect of the homogeneous solutions for $f$ and $\Phi_{J}$ in the BPS flow equations \eqref{steep}--\eqref{metflow}. For simplicity we will set $\beta =\varphi_2=\varphi_3=\varphi_4= 0$. Note that in the limit $a \to \infty$ we get the corresponding BPS equations in the Poincar\'{e} patch.

The BPS naked singularity at a fixed point for the scalars is given by \eqref{BPSnaked1}--\eqref{BPSnaked4}, where $A_0$ and $C_1$ are two integration constants which we can fix at will. We will find a domain wall which interpolates between the two naked singularities - the maximally symmetric one for which all scalars vanish and the PW point \eqref{critvals1}.  To find such a domain wall we will use equations \eqref{BPSnaked1}--\eqref{BPSnaked4} as the boundary conditions and will integrate \eqref{steep}--\eqref{metflow} numerically . We find it convenient to solve the flow equations as a function of $A$.

\begin{figure}[!ht]
\begin{center}
\includegraphics[width=8.5cm]{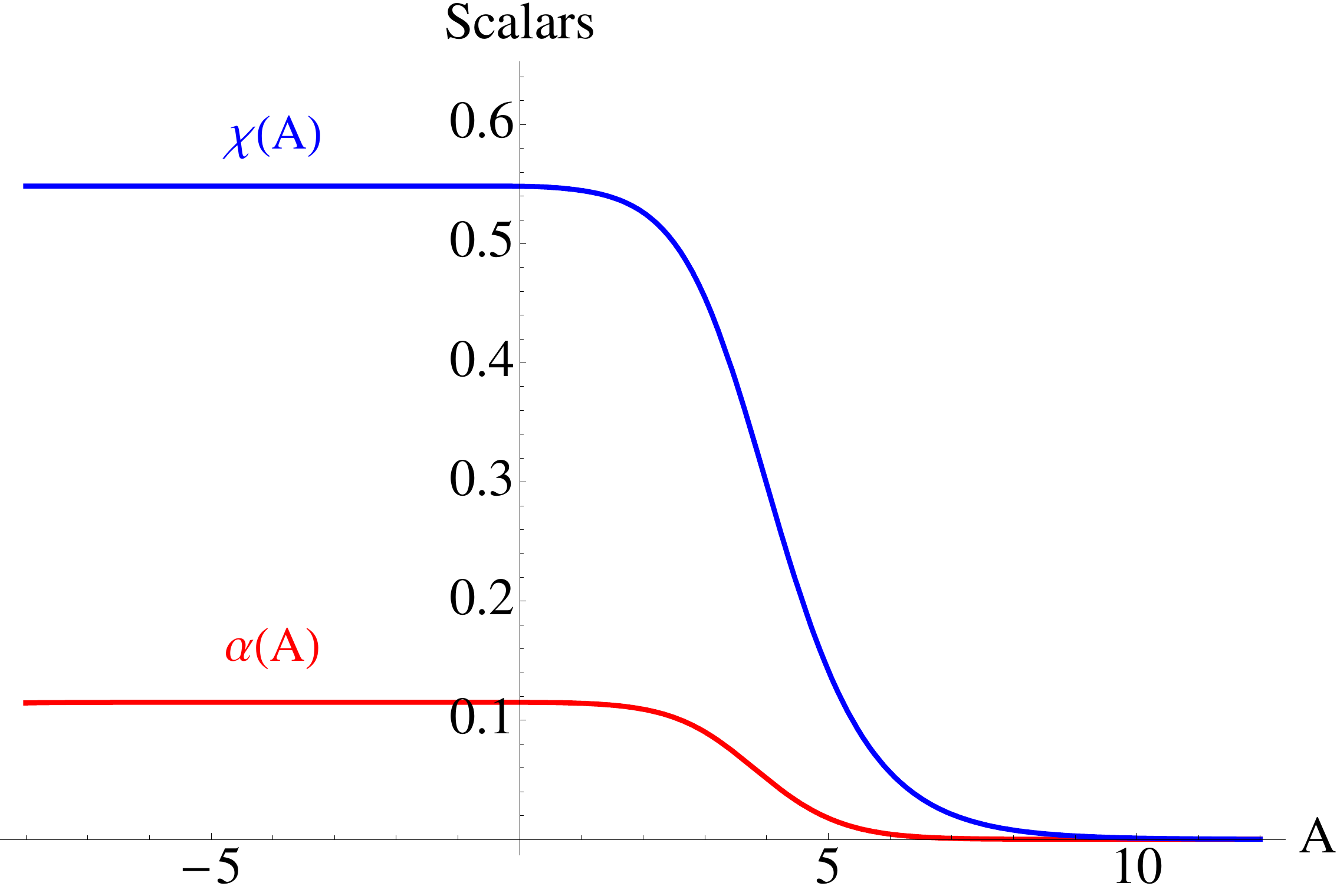}
\caption{{\it The profiles of the scalar functions $\alpha(A)$, $\chi(A)$ obtained by solving the BPS flow equations in Poincar\'e patch. We have also set $c = 1$, $g  = 1$. They do interpolate between the UV and the IR fixed points.}}
\label{figp}
\end{center}
\end{figure}

\begin{figure}[!ht]
\begin{center}
\includegraphics[width=8.5cm]{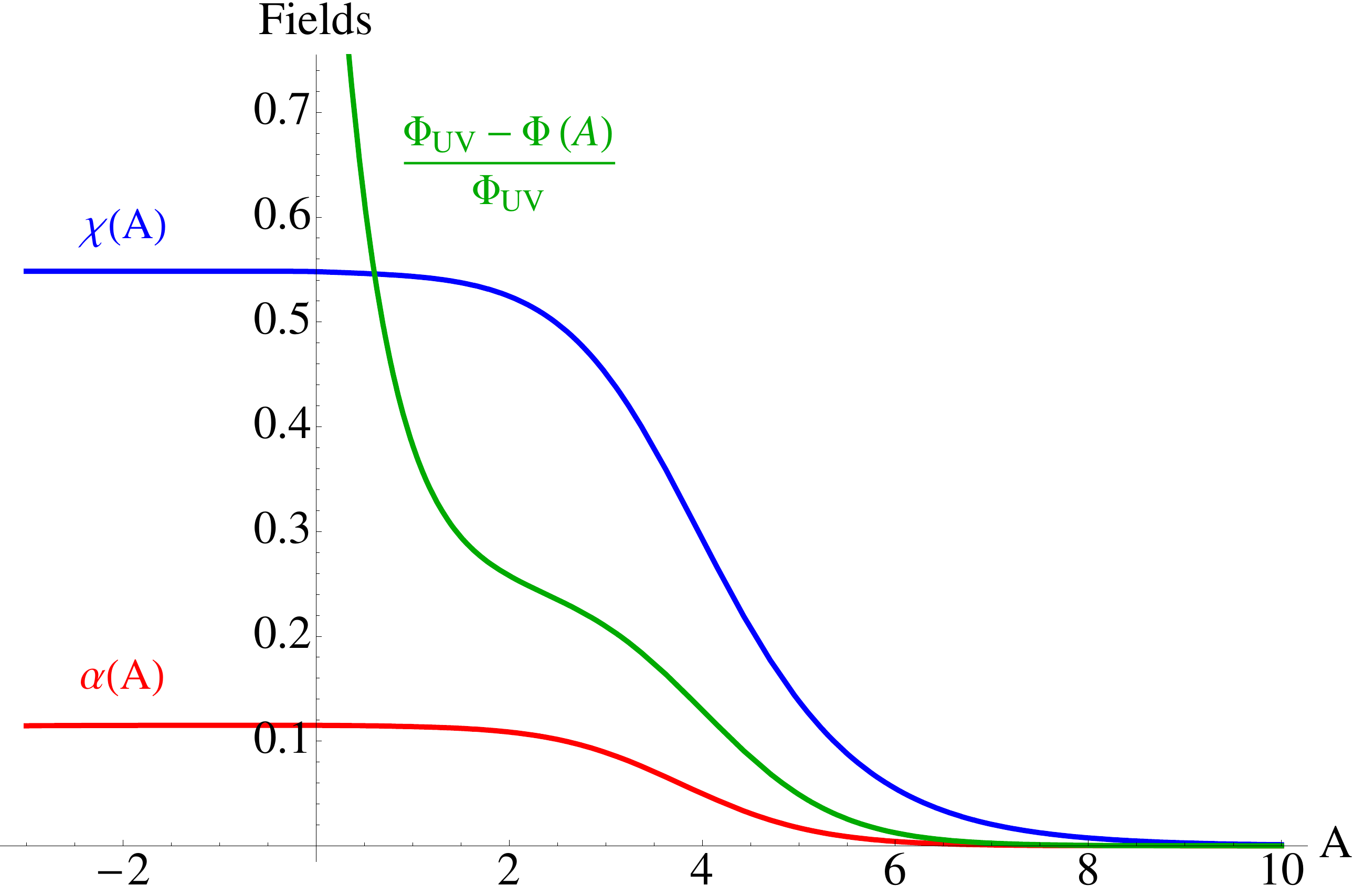}
\caption{{\it The profiles of the scalar functions $\alpha(A)$, $\chi(A)$ and $\Phi(A) \equiv \Phi_2(A)$ obtained by solving the BPS flow equations in global coordinates. For numerical purpose we have set $c = 1$, $g  = 1$ and $a = 1$. Note that all fields vary within the same region in $A$-space and that $\Phi$ is diverging exponentially and $f$ is determined by \eqref{Phians1}.}}
\label{figg}
\end{center}
\end{figure}

The numerical solution\footnote{It is again possible to guess a fitting tangent hyperbolic function that will approximate the numerical solution very closely, as we have demonstrated earlier in (\ref{exactalpha}) and (\ref{exactchi}).} in the Poincar\'e patch is shown in Fig.~\ref{figp}. For solutions asymptotic to global $AdS_5$ we also have non-zero values of the gauge fields and we can numerically solve the flow equations. The results are shown in Fig.~\ref{figg}. The linearized analysis around the end points of the flow is quite similar to the one done in Section 4 and there is again exponential damping in the linearized equation for the $\chi$-field as observed in (\ref{IRdiffeqns}). Indeed the scalars around the fixed points behave in the same way as in Section 4 and the behavior of the $A(r)$, $f(r)$ and $\Phi_I(r)$ is given by \eqref{BPSnaked1}--\eqref{BPSnaked4}. In particular $\Phi_I$ and $f^2$ are diverging, $\Phi_I \sim e^{ - 2 r / L_*} $ and $f^2 \sim e^{ - 6 r / L_*}$, as $r\to -\infty$.

%%%%%%%%%%%%%%%
%%%%%%%%%%%%%%%

%%%%%%%%%%%%%%%
%%%%%%%%%%%%%%%

It should be very interesting to study similar domain walls at finite temperature. Since such solutions would break supersymmetry one would have to solve second order flow equations derived from the equations of motion. Such analysis is beyond the scope of the current paper. In the following section we will study some simple non-supersymmetric black hole solutions.

%%%%%%%%%%%%%%%%%
\subsection{Non-supersymmetric solutions}
%%%%%%%%%%%%%%%%%

Even though, from the ten-dimensional perspective, the solutions constructed above have an interesting interpretation in terms of spinning D3-branes and giant gravitons, they are singular in five dimensions.  On the other hand, there are well-known, non-supersymmetric black-hole solutions that are regular outside their horizons and  thus it is interesting to see how such non-BPS solutions appear in the $\Neql2$ theory coupled to a charged hypermultiplet. 

One can set $\varphi_2=\varphi_3=\varphi_4=0$ and $\varphi_1$ to one of the two fixed point values and then derive the following equations of motion from the supergravity Lagrangian \eqref{Lagrangian}\footnote{Apart from the constraints obtained from setting both $\alpha$ and $\beta$ to their respective constant values, there is an additional constraint equation at the PW fixed point resulting from the variation of $\theta_1$.  It is straightforward to check that all these constraints imply the relation in (\ref{pwacon}).}:
\begin{eqnarray}
&& R_{\mu\nu} - g_{\mu\nu}\ds\frac{R}{2} - 2 \mathcal{P} g_{\mu\nu} + 2 (\rho^{4}\nu^{-4} T_{\mu\nu}^{(1)} + \rho^{4}\nu^{4} T_{\mu\nu}^{(2)} + \rho^{-8} T_{\mu\nu}^{(3)}) - 4 (3 T(\alpha)_{\mu\nu}+T(\beta)_{\mu\nu}) = 0~, \notag\\\notag \\
&&  \partial_{\sigma} \left(\sqrt{g} \rho^{4}\nu^{-4} F^{(1)\sigma}_{\nu} \right) = 0~, \quad \partial_{\sigma} \left(\sqrt{g} \rho^{4}\nu^{4} F^{(2)\sigma}_{\nu} \right) = 0~, \quad \partial_{\sigma} \left(\sqrt{g} \rho^{-8} F^{(3)\sigma}_{\nu} \right) = 0~, \qquad  \notag\\\\
&& \ds\frac{1}{\sqrt{g}} \partial_{\nu} \left(\sqrt{g} \partial^{\nu} \alpha \right) + \ds\frac{1}{6} \left(\rho^{4}\nu^{-4} F^{(1)}_{\mu\nu}F^{(1) \mu\nu} + \rho^{4}\nu^{4} F^{(2)}_{\mu\nu}F^{(2)\mu\nu} - 2\rho^{-8} F^{(3)}_{\mu\nu}F^{(3)\mu\nu}  \right) + \ds\frac{1}{6} \ds\frac{\partial \mathcal{P}}{\partial \alpha} = 0~, \notag \\\notag\\
&& \ds\frac{1}{\sqrt{g}} \partial_{\nu} \left(\sqrt{g}\partial^{\nu} \beta \right) + \ds\frac{1}{2} \left( - \rho^{4}\nu^{-4} F^{(1)}_{\mu\nu}F^{(1) \mu\nu} + \rho^{4}\nu^{4} F^{(2)}_{\mu\nu}F^{(2)\mu\nu}  \right) + \ds\frac{1}{2} \ds\frac{\partial \mathcal{P}}{\partial \beta} = 0~, \notag
\end{eqnarray}  
where
\begin{equation}
T_{\mu\nu}^{(i)} = F^{(i)}_{\mu\sigma}F^{(i)\sigma}_{\nu} - \coeff{1}{4}\,g_{\mu\nu} F^{(i)}_{\sigma\lambda}F^{(i)\sigma\lambda}~, \qquad T(X)_{\mu\nu} = \partial_{\mu} X\partial_{\nu} X - \coeff{1}{2}\,g_{\mu\nu} \partial_{\sigma}X \partial^{\sigma}X~,
\end{equation}
and $\mathcal{P}$ is the potential \eqref{PsuperP}. A solution to this set of equations is given by
\begin{equation}
ds^2_5 = \ds\frac{F}{(H_1H_2H_3)^{2/3}} dt^2 -  \coeff{1}{4} a^ 2 (H_1H_2H_3)^{1/3} \eta^2 (\sigma_1^2 + \sigma_2^2 + \sigma_3^2)  - \ds\frac{(H_1H_2H_3)^{1/3}}{F} d\eta^2~,
\end{equation}
\begin{equation}
H_{i} (\eta) = \ds\frac{g}{2c} \left(1 + \ds\frac{q_i}{\eta^2}\right)~, \qquad\qquad F(\eta) = \ds\frac{1}{a^2} - \ds\frac{M}{\eta^2}  + c^2 \eta^2 H_1H_2H_3 \,, 
\end{equation}
\begin{equation}
A_{(i)} = - \left(\ds\frac{M}{4q_i} + \ds\frac{1}{4a^2}\right)^{1/2} \ds\frac{1}{H_i} dt~,\qquad\qquad X^{(i)} =  \ds\frac{(H_1H_2H_3)^{1/3}}{H_i}~,
\end{equation}
This is the non-supersymmetric $AdS_5$ black hole of the STU model with three different charges \cite{Behrndt:1998jd}. The non-extremality parameter is $M$ and for $M=0$ we recover the BPS solution of Section \ref{STUBPS}. There is a critical value of $M$ above which the solution has no naked singularities and has a macroscopic horizon \cite{Behrndt:1998jd}.

One can also construct a non-BPS black hole with a constant charged scalar which is asymptotic to the PW fixed point\footnote{There is also a similar non-supersymmetric solution asymptotic to the $SU(3)$ fixed point \eqref{SU3point}.} \eqref{critvals1}. The solution is the non-supersymmetric generalization of the BPS naked singularity presented in Section \ref{PWBHBPS} and is given by    
\begin{equation}
ds^2_5 = \ds \widetilde{F} H^{-2} d\tau^2 -  \coeff{1}{4} a^ 2 H \eta^2 (\sigma_1^2 + \sigma_2^2 + \sigma_3^2) - \ds H\widetilde{F}^{-1} d\eta^2~,
\end{equation}
with 
\begin{eqnarray}
H(\eta) & = & \ds\frac{g}{2c} \left( 1 + \ds\frac{q}{\eta^2} \right) \ , \quad \widetilde{F}(\eta) = \ds\frac{1}{a^2} - \ds\frac{M}{\eta^2} + \ds\frac{4 c^2  W_{*}^2 }{9} \eta^2 H^3(\eta)~. 
\end{eqnarray}
The vector potentials are given by:
\begin{equation}
\begin{split}
A_{(1)} &= - \left(\ds\frac{M}{4q} + \ds\frac{1}{4a^2}\right)^{1/2} \ds\frac{\nu_{*}^2}{\rho_{*}^2}\,  \ds\frac{d\tau}{H}~,\qquad A_{(2)} =  - \left(\ds\frac{M}{4q} + \ds\frac{1}{4a^2}\right)^{1/2}\ds\frac{1}{(\rho_* \nu_{*})^2}\ds\frac{d\tau}{H}~, \\[7pt] 
A_{(3)} &= - \left(\ds\frac{M}{4q} + \ds\frac{1}{4a^2}\right)^{1/2}\rho_{*}^4\, \ds\frac{d\tau}{H}~,
\end{split}
\end{equation}
and the scalars have the constant values \eqref{constSTUscalars}. Once again, while the solution has the same form as a standard STU black hole, it does, of course, represent a new supergravity solution that possesses a very distinct, ten-dimensional uplift.
  
%%%%%%%%%%%%%%%%%%%%%%%%%%%%%%%%%%%%%%%%%%%%%%%%%%%%%%%%%%%%%%%%%
\section{Conclusions}
%%%%%%%%%%%%%%%%%%%%%%%%%%%%%%%%%%%%%%%%%%%%%%%%%%%%%%%%%%%%%%%%%

In this paper we have exhibited a large class of supersymmetric holographic flow solutions that involve chemical potentials and in which the metric on the brane is either that of $\IR^3 \times \IR^1$ or  $S^3 \times \IR^1$, with different warp factors in front of the temporal and spatial directions.  These flows are completely determined by a first order system that involves a modified steepest descent  on the superpotential.  The solutions include an interesting new smooth flow between two global $AdS$ backgrounds with different radii and different potentials.  The potential changes because there is a ``charged hypermultiplet cloud'' between the UV and IR fixed points.
 
Even within the families of solutions considered here, there are several other examples that would be interesting to examine, particularly the $\Neql{2}^*$ flows \cite{Pilch:2000ue},  the $\Neql{1}^*$ flow \cite{Girardello:1999bd,Pilch:2000fu} with three masses equal and hybrids of the two \cite{Evans:2000ap}. Then there are all the M-theory analogues of these flows.  One would expect that the flow to the $SU(3) \times U(1)$ invariant fixed point \cite{Warner:1983vz, Nicolai:1985hs, Ahn:2000aq, Ahn:2000mf, Corrado:2001nv} could be easily promoted to a smooth flow with chemical potentials in global $AdS$ in a manner very analogous to the result here \cite{progress}.  The flow to the supersymmetric $G_2$-invariant point is somewhat more problematic because there are no $G_2$-invariant vector fields in $SO(8)$ to provide suitable chemical potentials.  There might, however, be something far more exotic involving chemical potentials in non-trivial representations of the unbroken gauge group.  It might also be rather interesting to see what happens to the $SU(3)$-invariant web of flows considered in 
\cite{Bobev:2009ms} once one includes the chemical potentials.

As we remarked earlier, the gauge potentials that give rise to the chemical potentials correspond to Kaluza-Klein angular momentum terms on the $S^5$ of the compactification of IIB supergravity.  Thus the  uplifts of the solutions considered here, and their M-theory analogues, all correspond to some form of giant graviton solution.  It would be interesting to examine these uplifts in more detail and we have already begun work on this \cite{progress}.  Obviously the smooth solution would be most interesting, but even the singular black-hole-like solutions should have a much more canonical description in terms of spinning branes.  

Going beyond what we have considered here, it should be relatively straightforward to include magnetic fields in addition to the electrostatic potentials.  The inclusion of magnetic fields would not only enable us to study the dual AdS/CMT systems in the presence of magnetic fields and perhaps find new phenomena, like new phases, but would also give us a new approach to studying   $AdS$ black-hole solutions.   In particular, one should certainly recover the supersymmetric black hole in global $AdS$ \cite{Gutowski:2004ez,Gutowski:2004yv}, and perhaps put such a solution in the core of a cloud of charged hypermultiplets.     It would also be interesting   to use the approach presented here to re-examine, and possibly generalize, the solutions in \cite{D'Hoker:2009mm,D'Hoker:2009bc}.  More generally, one could also consider electromagnetic backgrounds that have ``instanton-like'' boundary conditions at infinity with an $SU(2)$ electromagnetic field winding non-trivially around the $S^3$ of global $AdS$. 

Based upon the study of black holes and black rings in asymptotically flat five-dimensional space (see, for eaxmple, \cite{Gauntlett:2004wh, Bena:2004de,Elvang:2004ds, Bena:2007kg}),  one expects that the supersymmetry equations will no longer be sufficient to determine the solution with both electric and magnetic fields and that one will have to supplement these equations with some of the equations of motion in order to fully determine the solutions.  The situation should be simpler with M-theory compactifications to four-dimensions because the magnetic fields are necessarily constant if they are uniform in space  and so should not require the use of the equations of motion in addition to the requirements of supersymmetry.  

Finally, there are the non-supersymmetric solutions.  Based upon our work here, it is evident that the supersymmetric flow between the fixed points necessarily involves turning on a fermion mass and not a fermion condensate.  On the other hand, there are non-supersymmetric flows involving a fermion condensate between these fixed points \cite{GPRprivate} and so there are clearly interesting non-supersymmetric flows to be studied.  Given that the end-points of such a non-supersymmetric flow are stable, it would be most interesting to learn whether the entire flow is stable.

It is also evident from our work that one can find solutions with non-supersymmetric black holes in the center of the hypermultiplet charge clouds.  It would be interesting to study flows that start from the maximally symmetric fixed point and evolve to such black-hole solutions in the core.  More generally, it may be that the study of $AdS$ black holes and black rings could be fertile territory for the application of ``almost-BPS'' techniques  \cite{Goldstein:2008fq} and their generalizations    \cite{Bena:2009ev, Bena:2009fi}.  The basic idea is to find solutions that are ``locally supersymmetric'' but for which global holonomies break the supersymmetry.  One might therefore hope to use this perspective to take many of the the supersymmetric black holes, black rings and bubbled solutions and carry them over to ``almost-BPS'' solutions in $AdS$ geometries.

The study of supersymmetric flow solutions involving background electromagnetic fields raises some extremely interesting questions and suggests quite a number of new avenues for further research.

%%%%%%%%%%%%%%%%%%%%%%%%%%%%%%%%%%%%%
\bigskip
\bigskip
\bigskip
\leftline{\bf Acknowledgements}
\smallskip
We would like to thank Steve Gubser, Sliviu Pufu and Fabio Rocha for stimulating our interest in holographic flows with chemical potentials.  We are also grateful to  Igor Klebanov and Edward Witten for very helpful conversations about holography on global $AdS$ spaces. NB would like to thank Tameem Albash for useful discussions on charged black holes. This work was supported in part by DOE grant DE-FG03-84ER-40168.  
%%%%%%%%%%%%%%%%%%%%%%%%%%%%%%%%%%%%%

%%%%%%%%%%%%%%%%%%%%%%%%%%%%%%%%%%%%%

%%%%%%%%%%%%%%%%%%%%%%%%%%%%%%%%%%%%%

\end{document}